\documentclass[twocolumn]{aastex63}

\usepackage{textcomp}
\usepackage{threeparttable}
\usepackage{booktabs} 

\newcommand{\dsct}{$\delta$~Sct~}

\newenvironment{Contfigure}{%
\renewcommand{\thefigure}{\arabic{figure} (Continued)}%
\addtocounter{figure}{-1}%
\begin{figure*}}{%
\renewcommand{\thefigure}{\arabic{figure}}%
\end{figure*}}

\received{September 29, 2020}
\revised{December 17, 2020}
\accepted{December 18, 2020}

\submitjournal{AJ}

\shorttitle{Variability in NGC~419}
\shortauthors{Mart\'inez-V\'azquez et al.}

\graphicspath{{./}{figures/}}
 
\begin{document}

\title{Short period variability in the globular cluster NGC 419 and the SMC field
\footnote{Based on observations obtained at the international Gemini Observatory, a program of NOIRLab, which is managed by the Association of Universities for Research in Astronomy (AURA) under a cooperative agreement with the National Science Foundation. on behalf of the Gemini Observatory partnership: the National Science Foundation (United States), National Research Council (Canada), Agencia Nacional de Investigaci\'{o}n y Desarrollo (Chile), Ministerio de Ciencia, Tecnolog\'{i}a e Innovaci\'{o}n (Argentina), Minist\'{e}rio da Ci\^{e}ncia, Tecnologia, Inova\c{c}\~{o}es e Comunica\c{c}\~{o}es (Brazil), and Korea Astronomy and Space Science Institute (Republic of Korea).}}

\correspondingauthor{C. E. Mart\'{i}nez-V\'{a}zquez}
\email{clara.marvaz@gmail.com}

\author[0000-0002-9144-7726]{C.~E.~Mart{\'\i}nez-V\'azquez}
\affiliation{Cerro Tololo Inter-American Observatory/NSF’s NOIRLab, Casilla 603, La Serena, Chile}

\author[0000-0002-1206-1930]{R.~Salinas}
\affiliation{Gemini Observatory/NSF’s NOIRLab, Casilla 603, La Serena, Chile}

\author[0000-0003-4341-6172]{A.~K. Vivas}
\affiliation{Cerro Tololo Inter-American Observatory/NSF’s NOIRLab, Casilla 603, La Serena, Chile}

\begin{abstract}
Delta Scuti ($\delta$~Sct) stars have been extensively studied in our Galaxy, but far less in extragalactic systems. Here we study the population of \dsct variables in NGC~419, an intermediate-age globular cluster of the Small Magellanic Cloud (SMC), using $g,r,i$ Gemini-S/GMOS time series observations. Our goal is to study the role of such variables in the cluster extended main-sequence turnoff (MSTO). We report the discovery of 54 \dsct stars and three eclipsing binaries in the NGC~419 field.
We find only a handful of the \dsct stars at the MSTO of NGC~419 while the majority is fainter, indicating that the cluster is younger ($\lesssim 1.2$~Gyr) than previously thought. Considering their radial distribution, we identify only six \dsct stars as probable members of NGC~419 while the 48 remaining are likely \dsct stars of the SMC field. Cluster \dsct stars appear close to the red edge of the MSTO, supporting the idea that the extended MSTO has its origin in an age spread. The 48 field \dsct stars represent the largest detection of \dsct stars made in the SMC. The period distribution of these newly detected \dsct stars ($0.04 \lesssim P \lesssim 0.15$~d) is similar to that detected in other systems. The amplitude distribution ($0.05 \lesssim \Delta r \lesssim 0.60$~mag) is likely biased because of the lack of low-amplitude stars.
We finally use the \dsct stars to calculate distances using different period-luminosity relations. The average distance moduli obtained are $18.76\pm0.14$~mag for NGC~419 and $18.86\pm0.11$~mag for the SMC field, which agree with previous measurements.
\end{abstract}


\section{Introduction} \label{sec:intro}

The variable star population in star clusters provide unique samples of variables in different evolutionary stages, but sharing the same distance, and chemical composition, therefore presenting tight constraints for stellar pulsation theories and the calibration of period-luminosity relations.

Even though variability in star clusters in the Milky Way have been extensively studied \citep[e.g.][]{clement01, salinas16a}, in the~Magellanic Clouds (MCs), the large variability surveys like OGLE \citep{udalski15} and SuperMACHO \citep{rest05} have only scratched the surface of variability in dense star clusters given the difficulties posed by crowding at these distances; and very few clusters have been targeted individually for variability outside these surveys \citep[e.g.][]{kuehn13}. In particular, the population of variable stars near the main sequence turn-off (MSTO) and below have remained completely unexplored until the recent study by \citet{salinas18} in the Large~Magellanic Cloud (LMC) cluster NGC 1846.

In old globular clusters we can find blue stragglers within the instability strip which pulsate as SX Phoenicis stars. On the other hand, in young and intermediate age systems, main sequence stars are the ones that cross the instability strip producing pulsating stars known as delta Scuti ($\delta$~Sct) stars. From the observational point of view it is very hard to separate both types of stars \citep[e.g.,][]{cohen12}. Thus, hereinafter, we are adopting the term \dsct stars to refer to all pulsating variables in this zone of the main sequence.
$\delta$~Sct stars are H-shell and H-core burning, pulsating variable stars located in the intersection of the main sequence and the instability strip. Their periods are found to be in the interval between 0.008--0.42~d and their amplitudes between 0.001--1.7~mag in the $V$-band \citep[e.g.,][]{catelan15}. It was not until the past decades that large samples of \dsct stars have become available \citep[e.g.,][]{rodriguez00, poleski10, garg10, vivas13, murphy19, jayasinghe20}. Several factors play a fundamental role in this kind of studies, being deep photometry and high cadence the most important ones. Particularly, the recent studies of \citet{murphy19} and \citet{jayasinghe20} detect several thousands of \dsct stars in the Galactic field from the \textit{Kepler} and ASAS-SN catalogs, respectively.

This kind of variable stars, near the MSTO, may also play an unexpected role in the case of intermediate-age (1--3~Gyr) clusters in the MCs, which show extended or split MSTOs \citep[e.g.][]{mackey07,milone09}.

Two main mechanisms to produce these broad MSTOs have been mainly studied in the literature: an age difference, where either an extended or bursty star formation would produce sub-populations inside a star cluster whose different ages would be reflected in extended or split MSTOs \citep[e.g.][]{goudfrooij11}; or an spread in different stellar rotations \citep[e.g.][]{bastian09,brandt15}, without excluding the presence of both at the same time \citep{goudfrooij17,dupree17,gossage19}.

A third mechanism was proposed by \citet{salinas16b}, invoking that variable stars produced in the intersection of the instability strip and the main sequence (\dsct stars), will introduce a spurious broadening of the MSTO of intermediate-age clusters if not taken properly into account. This mechanism was tested with time series photometry in the LMC cluster NGC~1846 finding $\sim$50 \dsct stars, the largest number in any star cluster \citep{salinas18}. Though this number falls short by about an order of~magnitude to introduce a significant spread at the MSTO level of NGC 1846, the difficulties of studying stellar variability within the half-light radius of the cluster leaves the total number of \dsct stars still to be discovered, and the overall significance of these variables within the MSTO phenomenon in this and other clusters not fully addressed.

\subsection{The intermediate-age globular cluster NGC~419}

NGC~419 (RA = 01:08:17.8, Dec = -72:53:02.8) is an intermediate-age ($\sim 1.5$~Gyr), massive \citep[$\sim 1\times 10^5$ M$_{\odot}$,][]{kamann18,song19} globular cluster in the Small~magellanic Cloud (SMC). Having one of the first discovered extended MSTOs \citep{glatt08}, NGC~419 is a near twin to the LMC cluster NGC~1846, and therefore a good candidate to search for \dsct stars. The additional advantage of being tentatively located at around $\sim$ 10 kpc in front of the SMC body \citep{glatt08}, provides an environment where MSTO variability can be studied, in principle, with a minimized influence of the field population.

In this work we perform a deep (down to $g \sim 23$~mag) time-series photometry of NGC~419, the first star cluster in the SMC where short period variability is studied. Moreover, it offers an additional case to study the impact of variability at the MSTO morphology.

This paper is structured as follows. In \S~\ref{sec:obs} we present a summary of the observations and describe the data reduction, including the photometry, the variable star search, and determination of their pulsation parameters (i.e., periods, amplitudes, and intensity-averaged~magnitudes). In \S~\ref{sec:variables} we show the variable stars discovered and detected in this work and match them with previous detections made by OGLE and \textit{Gaia}. In \S~\ref{sec:dsct} we analyse the population of \dsct stars that may belong to NGC~419 and to the field of the SMC and we study the impact of variability at the MSTO morphology of NGC~419. In \S~\ref{sec:period-amplitude} we compare the period and amplitude distributions with other \dsct star catalogs available in the literature, and in \S~\ref{sec:distance} we determine distances from a new period-luminosity relation using the \dsct stars. Finally, in \S~\ref{sec:conclusions} we set out the main results and conclusions of this work.  

\section{Observations and data reduction} \label{sec:obs}

Time series imaging of NGC~419 was obtained with the Gemini Multi-Object Spectrograph (GMOS) mounted at the Gemini South telescope at Cerro Pach\'{o}n, Chile. Observations were carried out in the nights of September 14, 22 and 27, 2017, as part of the Fast Turnaround program GS-2017B-FT-2. Observations were obtained with 2$\times$2 binning, resulting in a pixel scale of 0.16 pixel arcsec$^{-1}$. Images were obtained in SDSS $g$, $r$ and $i$, with 11, 112 and 7 frames in total, with exposure times of 120, 90 and 80 seconds, respectively, reaching a S/N $\sim 50$ at the level of the MSTO. In order to build optimal light curves for \dsct stars stars in the $r$ band, we collected 35 epochs in the first night ($\sim 1.2$ consecutive hours), 64 epochs in the second night ($\sim 2.8$ consecutive hours) and 13 epochs in the last night ($\sim 0.5$ consecutive hours). We note that the $g$ and $i$ frames were interspersed with the $r$ frames in order to have the color information.

Data was reduced with the {\sc IRAF/Gemini} tasks. Specifically, all images were bias subtracted, flat field corrected and mosaiced. Additionally, unilluminated areas of the CCDs were trimmed. Image quality was measured with the {\sc gemseeing} task in {\sc IRAF}. $r$-band images FWHM varied between 0.46\arcsec and 1.18\arcsec (0.78\arcsec averaged), while $g$ and $i$ images had a mean FWHM of 0.92 and 0.72\arcsec, respectively.

\subsection{Photometry} \label{sec:photometry}

The photometry was performed using {\sc daophot iv/allstar} \citep{stetson87,stetson94}. An empirical point spread function (PSF) was derived for each image using $\sim$300 stars that were not saturated, far from the edges of the CCDs, and spread over the whole field-of-view (FOV) -- avoiding the central crowded region of the cluster. A Moffat function was the preferred PSF model over the whole set of images since it provided the smaller residual. Initially, a constant PSF photometry on individual images was obtained using {\sc allstar} which, in addition to the PSF photometric catalog, also provides a star-subtracted image. Because of the crowding in our field, some of the faintest sources remained undetected. We identified these faint sources using the star-subtracted image and appended them to our previous catalog in order to obtain a complete catalog for every image. Finally, subtracting the PSF stars from the individual images, we refined the PSF model and also allowed a quadratic variation of the PSF through the FOV in order to perform the final PSF photometry of our catalog.

Aperture corrections were measured based on a handful of isolated PSF stars, and finally, the photometry was calibrated into the standard system with observations of the SDSS standard field 020000-300600\footnote{https://www-star.fnal.gov/} obtained on the night of September 14, 2017, the first night of the observations. 

The color-magnitude diagram (CMD) from these GMOS observations is presented in Figure~\ref{fig:gmos_cmd_variables}. The CMD reaches $r = 24.5$ and $ g - r = 0.5$ with $\sigma = 0.05$~mag and $\sigma = 0.11$~mag, respectively. The photometric precision at the level of the MSTO ($r \simeq 20.5$) is $0.002$~mag in $r$ and $0.004$~mag in $g$.

\subsection{Variability search and period determination}

After testing with different variability indices \citep[see e.g.][]{sokolovsky17}, we used the robust median statistic (RoMS, \citealt{enoch03}) and the Welch-Stetson variability index \citep[J,][]{welch93,stetson96} as optimum variability indices to search for variable stars in the GMOS field centered on NGC~419. We visually inspected a few thousands light curves that were selected with RoMS $>$ 1 and J $>$ 2. The bulk of them were artifacts or background galaxies, recognized by a simultaneous visualization of the image and the raw time series data. In the same way, variable stars were easily recognized thanks to their light curve variation. 

Complementary, variables sources were also searched with the image subtraction program ISIS \citep{alard00}. ISIS first registers images to a common astrometric solution. A reference is constructed based on the best quality images, and then convolved to match the PSF of each image before subtraction. All subtracted images are stacked in a variance image, which is then visually inspected for residuals consistent with intrinsically varying sources \citep[e.g.][]{salinas16a}.

Periodograms were produced for each variable star candidate using Fourier analysis of the time series \citep{horne86}. Once periodicity was confirmed, the final period was refined by visually inspecting the light curves in the three bands simultaneously. The intensity-averaged magnitudes and amplitudes of the confirmed \dsct stars were derived by fitting the light curves with a set of templates (mainly based on the set of \citealt{layden98}). The best-fitting template was calculated using the \textit{Simplex} method \citep{nelder65}, minimizing the $\chi^2$ of the set of Layden's templates for a particular folded light curve. By obtaining the mean~magnitude and amplitude through the integration of the best-fitting template, we avoid biases appearing from light curves that are not uniformly sampled. This was particularly useful when calculating the mean~magnitude and amplitude for the light curves obtained in the $g$ and $i$ band, due to the small number of epochs. It is worth noting that Layden's templates were designed for RR~Lyrae stars. However, due to the lack of templates for \dsct stars and given the resemblance between the light curves of \dsct stars and those of RR~Lyrae stars, we used them to fit the \dsct light curves.

\newpage

\section{Variable stars}\label{sec:variables}

\begin{figure*}
    \centering
    \includegraphics[width=0.32\textwidth]{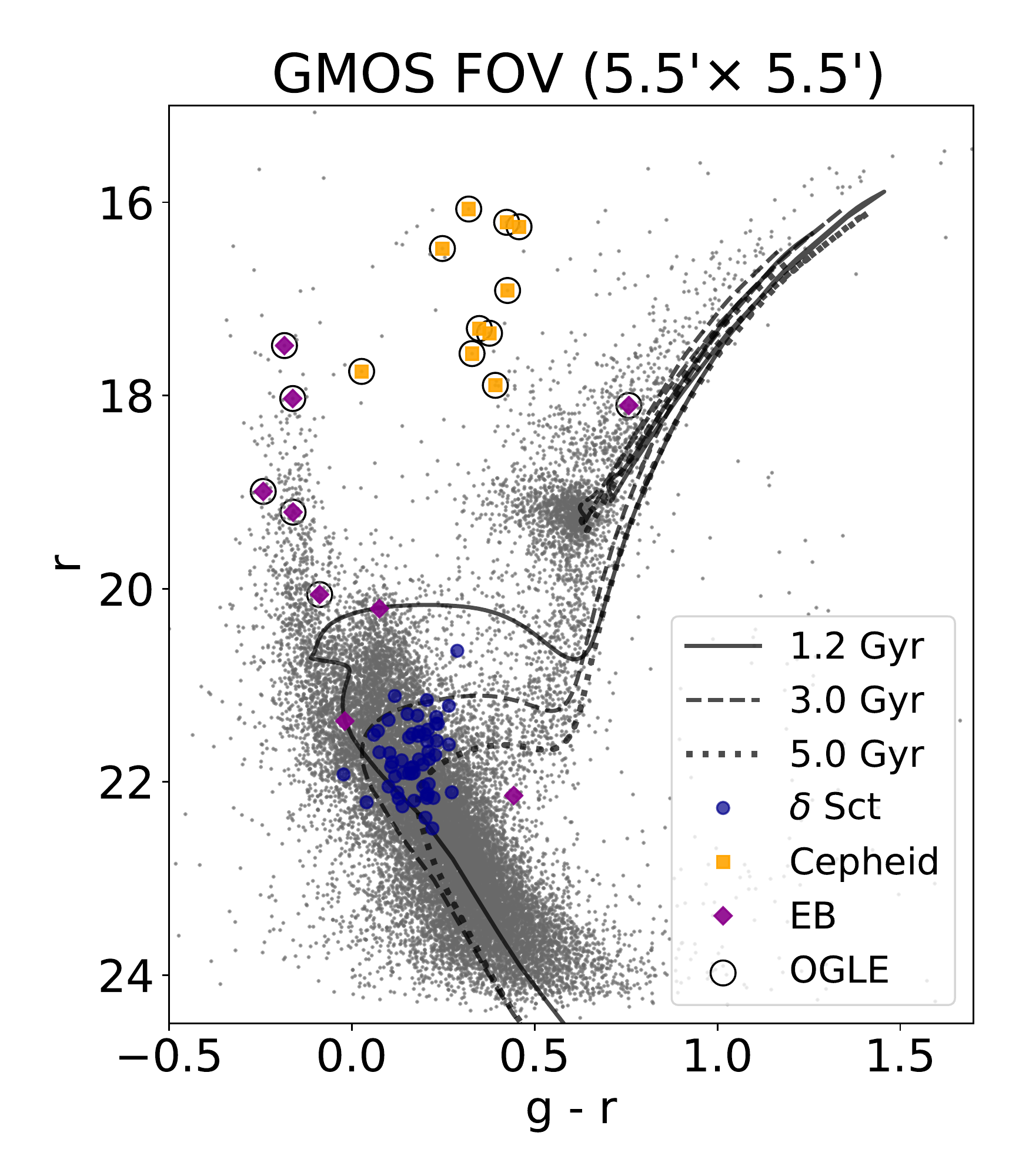}
    \includegraphics[width=0.32\textwidth]{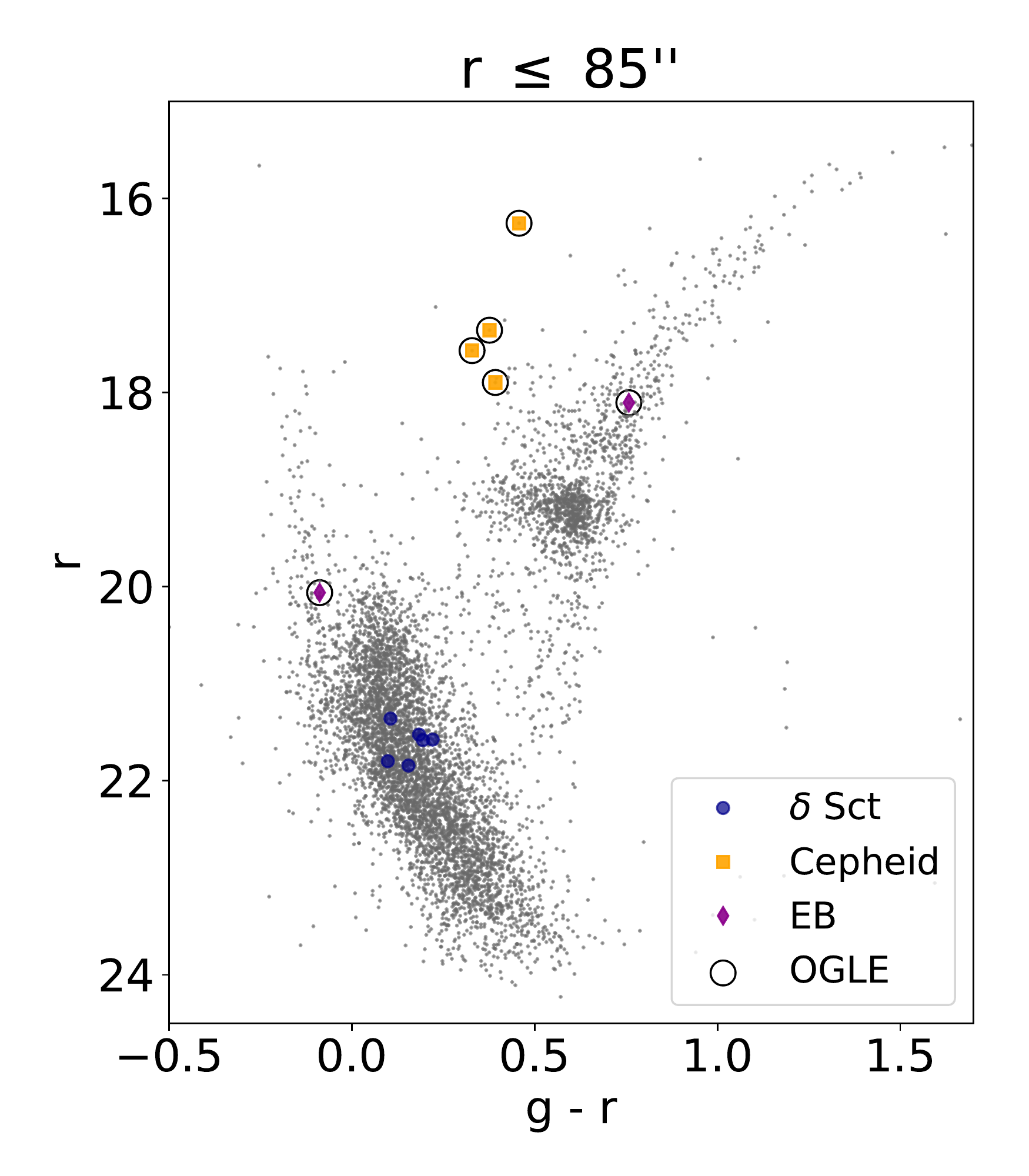}
    \includegraphics[width=0.32\textwidth]{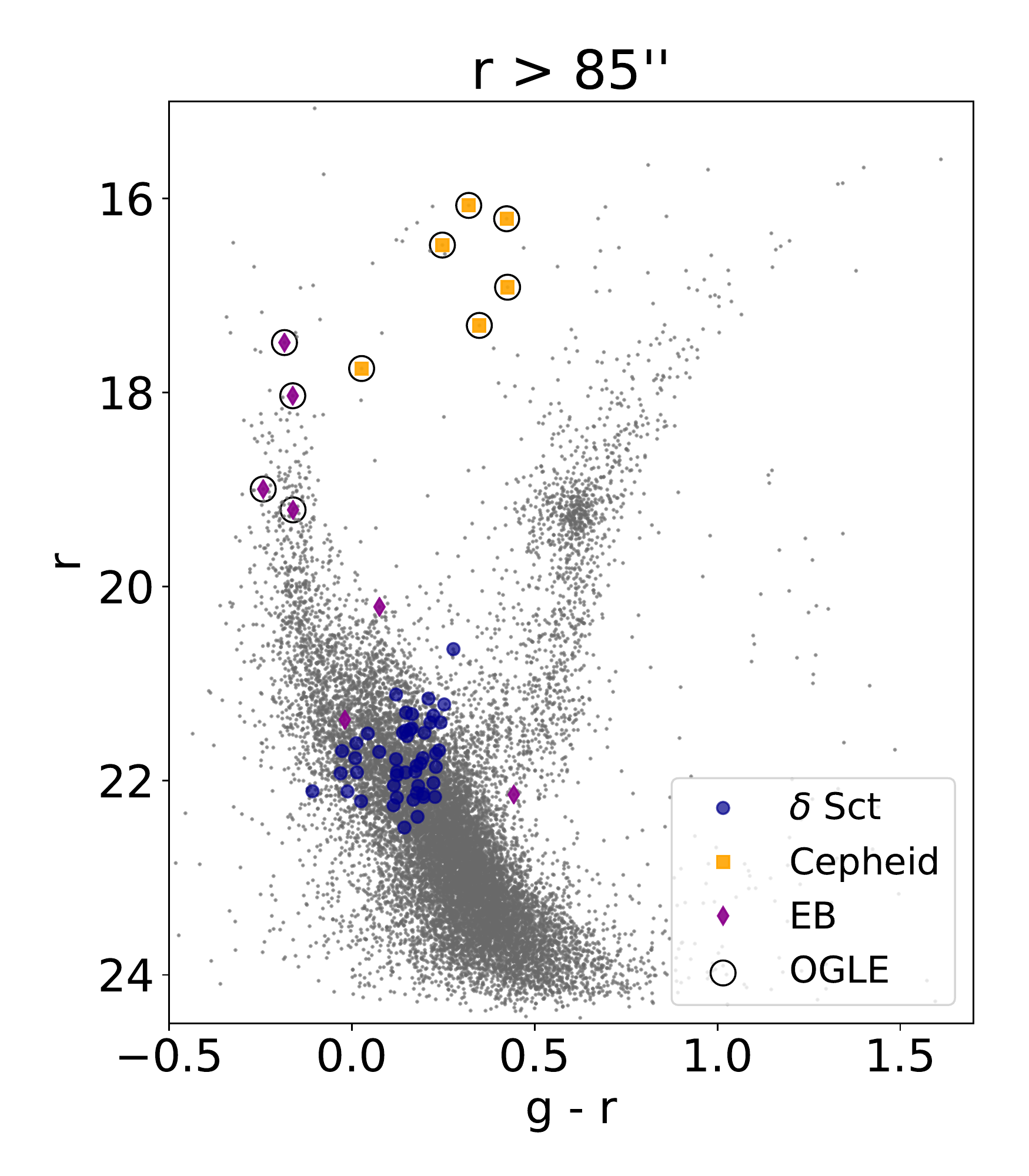}
    \caption{\textit{Left}.-- Color-magnitude diagram ($g-r$ vs $r$) of the stars within the GMOS FOV ($\sim$5.5\arcmin$\times$5.5\arcmin) centered on NGC~419. Variable stars detected and discovered in this work are shown with different symbols and colors. Blue circles represent the \dsct stars population, while orange squares and purple diamonds show the population of Cepheids and eclipsing binaries, respectively. Variables previously detected by OGLE are encircled. Black lines in the left panel represent the scaled-solar isochrones of 1.2~Gyr for $ \rm{[Fe/H]} = -0.55$~dex (cluster), and 3.0 and 5.0~Gyr for $ \rm{[Fe/H]} = -1.0$~dex (SMC field) obtained from BaSTI \citep{hidalgo18}. The isochrones were shifted assuming A$_V=0.25$~mag and $\mu_0=18.75$~mag for the 1.2~Gyr isochrone (cluster) and $\mu_0=18.97$~mag for the older isochrones (SMC field). Extinction coefficients were obtained as A$_{g} = 1.206~A_V$, A$_{r} = 0.871~A_V$ using \citep{cardelli89} and \citet{odonnell94} reddening law with R$_V=3.1$. \textit{Middle}.-- Color-magnitude diagram for the stars within 2~r$_h$ of NGC~419 ($\simeq 85$\arcsec). \textit{Right}.-- Color-magnitude diagram for the stars outside 2~r$_h$ of NGC~419.}
    \label{fig:gmos_cmd_variables}
\end{figure*}

\begin{figure*}
    \centering
    \includegraphics[width=0.75\textwidth]{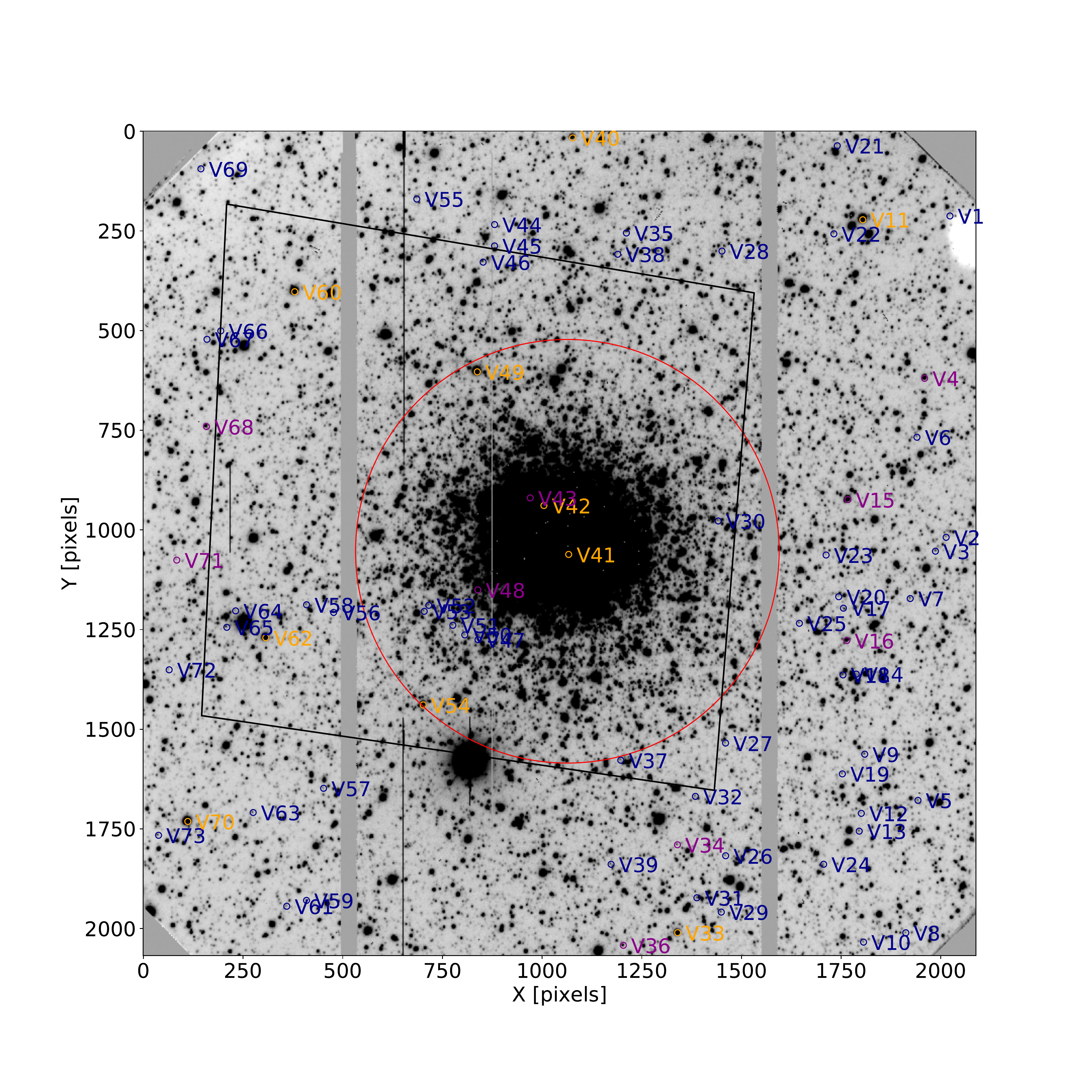}
    \caption{Finding chart of the variable stars detected inside the GMOS FOV ($\sim$5.5\arcmin$\times$5.5\arcmin) centered on  NGC~419. North is up and East is to the left. Blue circles represent the \dsct stars population, while orange and purple ones are the population of Cepheids and eclipsing binaries, respectively. The large black square displays the HST/ACS field observed in this cluster by \citet{milone09} and \citet{martocchia17}. The large red circle points the fitting radius of NGC~419 (85\arcsec, \citealt{ripepi14}).}
    \label{fig:finding_chart}
\end{figure*}

Table~\ref{tab:var} in Appendix~\ref{sec:puls-params} gives the positions, periods, intensity-weighted magnitudes and classifications for the 73 variable stars detected in this work. The individual $r,g,i$ measurements for the variables are published in Table~\ref{tab:photometry} in Appendix~\ref{sec:time-series}. Their folded light curves in $r$-band are presented in the Appendix~\ref{sec:lcv_dsct} and \ref{sec:lcv_ebin} (see Figures~\ref{fig:dSct_lcv} and \ref{fig:bin_lcv}, respectively), and Table~\ref{tab:comments} in Appendix~\ref{sec:comments} keeps record of individual notes for some of the variables. Out of them, 16 were previously detected by OGLE and 57 of them are new discoveries as we discuss in the following subsections. Figure~\ref{fig:gmos_cmd_variables} shows the location of these variables in the CMD. The variables already detected by OGLE are encircled. Furthermore, the spatial distribution (finding chart) of the variable stars detected is shown in Figure~\ref{fig:finding_chart}. 

\subsection{Known variables}

Using the online search tool provided by OGLE\footnote{\url{http://ogledb.astrouw.edu.pl/~ogle/OCVS/}}, a total of 17 OGLE variables were listed within our FOV \citep{soszynski15a,soszynski15b, soszynski16, pawlak16, soszynski19}. Sixteen of those were detected in our photometry and one fell into one of the CCD gaps of GMOS (the only SMC RR~Lyrae star in the field). Ten of them are classified as classical Cepheids by OGLE while the other six are eclipsing binaries. However, because of the age of NGC~419 ($\sim1.5$~Gyr), the cluster is too young to host any RR~Lyrae stars ($t>10$~Gyr, \citealt{walker89, savino20}) and too old to produce Cepheids ($t<300$~Myr, \citealt{bono05}). Therefore none of them are members of the cluster but of the SMC field instead. Due to our observation strategy, focused on short-period variables ($0.01 \lesssim$ P $\lesssim$ 0.15~d), periods were difficult to obtain for those variable stars with periods longer that $\sim$ 0.15~d, which are beyond the scope of this work. Therefore, the periods listed in Table~\ref{tab:var} for the OGLE variables, which are all longer than 0.15~d, are those published by OGLE.

Additionally, 11 stars classified as variables in \textit{Gaia} DR2 fall inside the GMOS FOV of NGC~419. Nine of them were detected in our photometry, eight are Cepheids already detected by OGLE (despite \textit{Gaia} classifying V40 as a RR~Lyrae star) and one is a long period variable (Mira) and because of that it is not listed in Table~\ref{tab:var}\footnote{It is very difficult to identify and be certain about the variation of long period variables such as Miras in our photometry. However, this star has a relatively high variability index  (RoMS = 4.5 and J = 5.7).}. The other two variable stars fall into the eastern gap of the CCD of GMOS. One of them corresponds to the OGLE RR~Lyrae star discussed previously, and the other one is a long period variable.

No \dsct stars have been discovered so far in this cluster.

\subsection{New variables}

In total, we have discovered in this work 57 variable stars: 54 \dsct stars and three eclipsing binaries. No additional Cepheids nor RR~Lyrae stars were discovered within the NGC~419 GMOS FOV. In particular, this is the first detection of \dsct stars in this cluster and in this region of the SMC field.  

The typical photometric uncertainty attained for the \dsct stars in each bands is $\sigma = 0.02$~mag. These errors come directly from the PSF fitting process with {\sc daophot/allstar} for each epoch. This considers the noise properties of the sources and detector, plus the accuracy of the PSF fitting.

\section{The population of \dsct stars in NGC~419 and the SMC field}\label{sec:dsct}

This work presents the first substantial detection of confirmed \dsct stars stars ever made in a field of the SMC. The left panel of Figure~\ref{fig:gmos_cmd_variables} shows the location of the \dsct stars (blue circles) in the CMD of the GMOS FOV centered on NGC~419, while the middle and right panels show the CMD within and outside 2~r$_h$ of NGC~419, respectively (see \S~\ref{sec:ngc419} for more details about this). Due to the similarities mentioned in \S~\ref{sec:intro} between NGC~419 and NGC~1846, it is salient that the bulk of the population of \dsct stars is not located in the MSTO of NGC~419 but $\sim$1~mag below it, which is in stark contrast with NGC~1846 where the \dsct stars population was detected mostly right at the MSTO of the cluster \citep[Figure 1 in][]{salinas18}. 

\begin{figure}
    \centering
    \includegraphics[width=1.0\columnwidth]{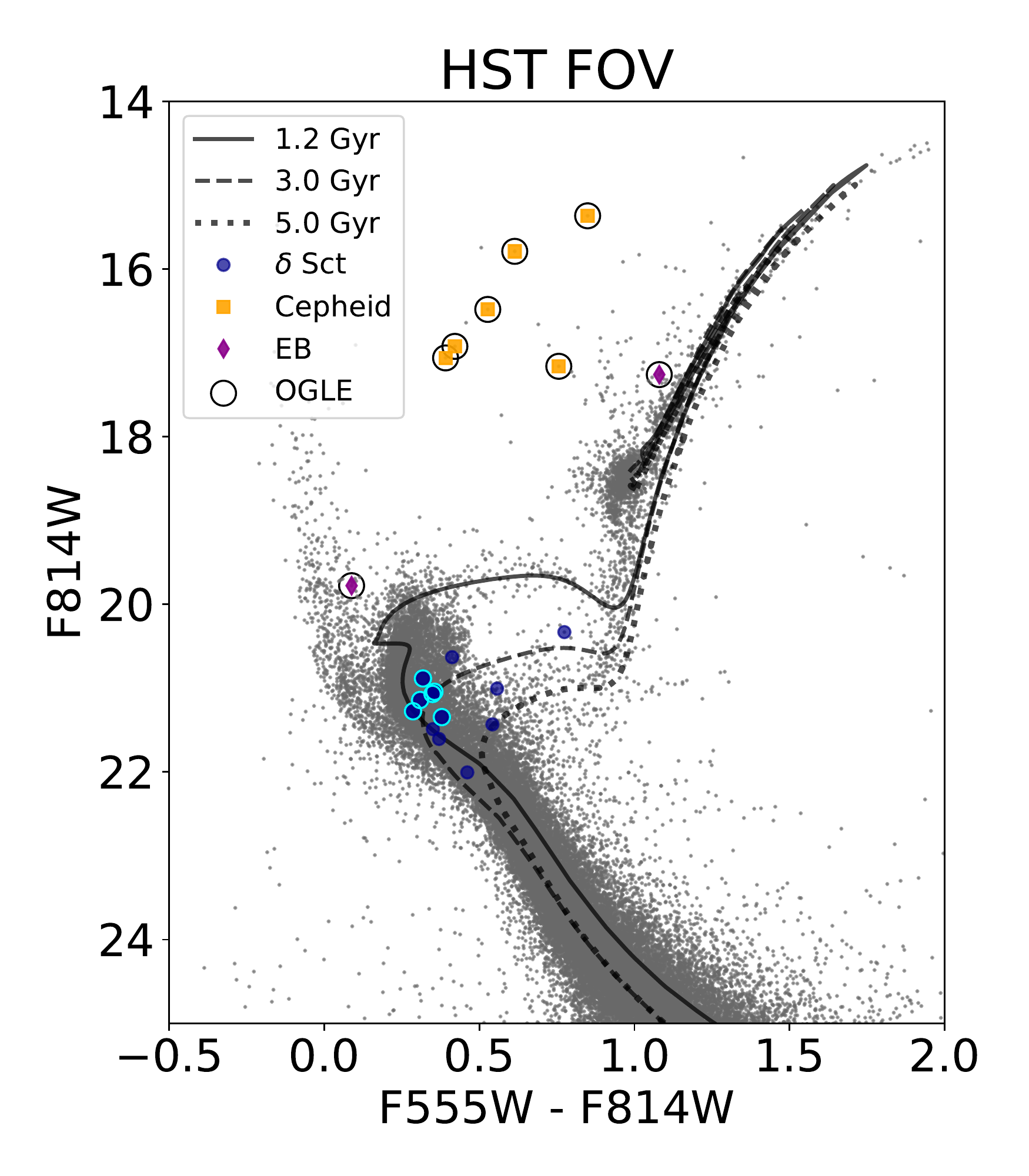}
    \caption{Color-magnitude diagram ($F555W-F814W$ vs $F814W$) of the \textit{HST}/ACS FOV pointing to NGC~419 from \citet{martocchia17}. Variable stars detected and discovered in this work are shown with different symbols and colors. Blue circles represent the \dsct stars population, while orange squares and purple diamonds show the population of Cepheids and eclipsing binaries, respectively. Variables detected previously by OGLE are encircled. The six \dsct stars within 2~r$_h$ of NGC~419 are highlighted with a cyan contour. Black lines represent the scaled-solar isochrones of 1.2~Gyr for $\rm{[Fe/H]} = -0.55$~dex (cluster), and 3.0 and 5.0~Gyr for $\rm{[Fe/H]} = -1.0$~dex (SMC field) obtained from BaSTI \citep{hidalgo18}. The isochrones were shifted assuming A$_V=0.25$~mag and $\mu_0=18.75$~mag for the 1.2~Gyr isochrone (cluster) and $\mu_0=18.97$~mag for the older isochrones (SMC field). Extinction coefficients were obtained as A$_{F555W} = 1.055~A_V$, A$_{F814W} = 0.586~A_V$ \citep{goudfrooij09, milone15}.}
    \label{fig:hst_cmd_variables}
\end{figure}

In order to understand the lack of \dsct stars in the MSTO of NGC~419, we use the \citet{martocchia17} \textit{HST}/ACS photometry of NGC~419 and identify the \dsct stars that are within that FOV. Figure~\ref{fig:hst_cmd_variables} shows the CMD of NGC~419 for the $F814W$ and $F555W$ bands. From this Figure, we notice that the \dsct stars are located in the red side of the MSTO of NGC~419. This may indicate that the cluster is younger than expected (1.5~Gyr) and that the blue side of the MSTO falls outside the instability strip, and therefore is not producing \dsct stars. Using the accurate \textit{HST}/ACS photometry, we fit isochrones to the NGC~419 and to the SMC field population to distinguish between the MSTO of the cluster and that of the SMC field to get some hint of the parent population of the \dsct stars (see Figure~\ref{fig:hst_cmd_variables}). The same isochrones are also displayed in the GMOS photometry (left panel of Figure~\ref{fig:gmos_cmd_variables}).

The reddening of NGC~419 is E($B-V$) $\simeq 0.08$~mag \citep{schlafly11} according to the NASA/IPAC Infrared Science Archive. This is consistent with recent values of the reddening based on Red Clump stars from OGLE-IV \citep{bell20}. Extinction coefficients were obtained as A$_{F555W} = 1.055~A_V$ and A$_{F814W} = 0.586~A_V$ \citep{goudfrooij09, milone15}, A$_g = 1.206~A_V$ and A$_r = 0.871~A_V$ (using \citet{cardelli89} and \citet{odonnell94} reddening law of R$_V =3.1$), where the $A_V$ of NGC~419 is 0.25~mag.

The isochrone overlaid on the CMD of NGC~419 is independent to those presented by \citet{glatt08}, \citet{girardi09} and \citet{goudfrooij14}. We fix the reddening and the metallicity parameters (which are known) and we adjust the age and the distance modulus ($\mu_0$) of the cluster. Most studies of the cluster have used a metallicity of $\rm{[Fe/H]} = -0.7$ which comes from a Calcium triplet measurement of \citet{kayser07}. However, here we employ a new measurement of the metallicity of the cluster --based on high resolution spectroscopy and measured directly from the Fe lines -- that indicates a slightly more metal rich cluster with $\rm{[Fe/H]} = -0.55$ (Alessio Mucciarelli, private communication and see \citealt{mucciarelli14}). With this metallicity we find -- using the updated BaSTI isochrone library \citep{hidalgo18}\footnote{http://basti-iac.oa-abruzzo.inaf.it/isocs.html} -- that the cluster is well described by a 1.2~Gyr population and a distance modulus of $\mu_0 = 18.75$~mag (see solid curve in Figure~\ref{fig:hst_cmd_variables}, also displayed in the left panel of Figure~\ref{fig:gmos_cmd_variables}). The values obtained in this fitting are different from those derived by \citet[][$t = 1.5$~Gyr and $\mu_0$=18.85~mag]{goudfrooij14}, but they are more similar to those derived by \citet[][particularly with their best-fitting isochrone of the Padova isochrone model, $t = 1.25$~Gyr and $\mu_0 = 18.75$~mag]{glatt08} and \citet[][$t = 1.35$~Gyr and $\mu_0 = 18.73$~mag]{girardi09}. 

For the field of the SMC, we have used two isochrone with ages 3 (dashed curve) and 5~Gyr (dotted curve), both with $\rm{[Fe/H]} = -1.0$ dex \citep{carrera08b, mucciarelli14}, encompassing the bulk of the SMC field population. These isochrones were shifted taking account of the reddening of NGC~419 and a distance modulus $\mu_0 = 18.97$~mag \citep{graczyk14}.

\begin{figure*}
    \centering
    \includegraphics[width=1.0\textwidth]{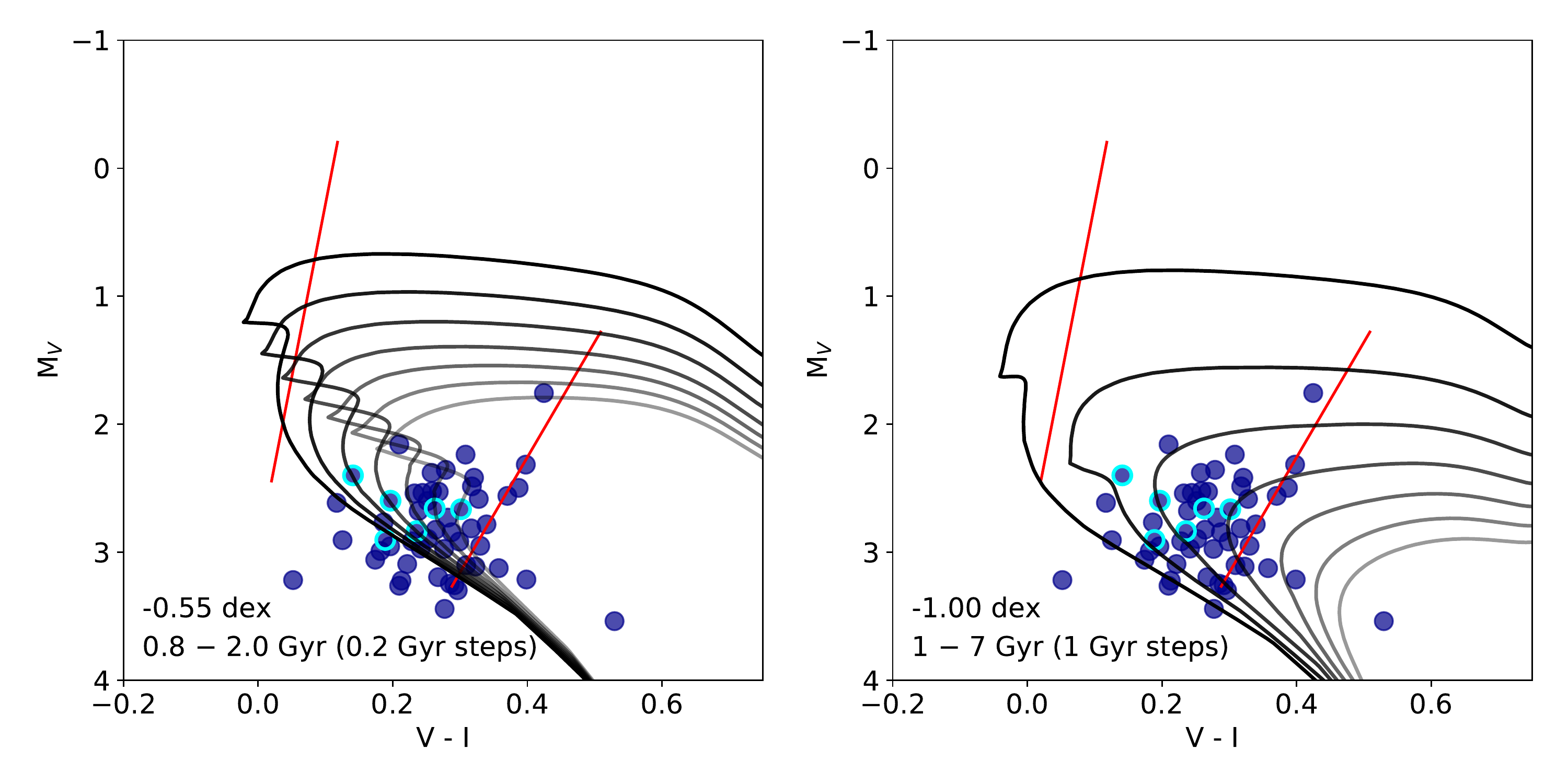}
    \caption{Color-magnitude diagram $V-I, V$ for the \dsct stars discovered in this work (blue circles). Red lines show the edges of the instability strip derived by \citet{murphy19} from \textit{Kepler} \dsct stars.
    Solid black lines in the left panel display the BaSTI isochrones \citep{hidalgo18} from 0.8 to 2~Gyr (with steps of 0.2~Gyr) for a $\rm{[Fe/H]} = -0.55$~dex, while in the right panel represent the isochrones from 1 to 7~Gyr for a $\rm{[Fe/H]} = -1.0$~dex (with steps of 1~Gyr). In both panels the gradient of colors goes from the youngest isochrone in black to the oldest one in light gray. The six \dsct stars within 2~r$_h$ of NGC~419 are highlighted with a cyan contour.}
    \label{fig:CMD_iso_IS}
\end{figure*}

To have more clarity on what population the progenitor of the discovered \dsct stars is, we show in Figure~\ref{fig:CMD_iso_IS} the \dsct stars detected in this work\footnote{$V$ and $I$ were obtained using the Lupton (2005) photometric transformation equations: \url{http://www.sdss3.org/dr8/algorithms/sdssUBVRITransform.php\#Lupton2005}. The absolute values were obtained after correcting for a distance modulus of 18.75~mag and A$_V=0.25$~mag.}, the instability strip and a set of isochrones with ages and metallicities consistent with the cluster and the field. The limits of the instability strip for Galactic \dsct stars, measured empirically by \citet{murphy19}\footnote{Since the edges of this instability strip were in $L$ and $T_{\rm eff}$, we convert them into $M_V$ and $V-I$ using the intrinsic colors, adopted $T_{\rm eff}$, and bolometric correction for A0V-F9V stars given by \citet{pecaut2013} in their Table 5.}, can be seen as solid red lines in Figure~\ref{fig:CMD_iso_IS}. A set of BaSTI isochrones with $\rm{[Fe/H]} = -0.55$~dex and ages from 800 Myr to 2~Gyr (in steps of 200 Myr) is displayed in the left panel of Figure~\ref{fig:CMD_iso_IS}. These encompass the range of ages that have been proposed for the cluster \citep{glatt08,rubele10,ripepi14}. From this Figure we note that the MSTO of a population of $\sim$1~Gyr would fall outside the instability strip and therefore no pulsations would be developed for stars in the blue edge of the NGC 419 MSTO. We stress that the edges of the instability strip, especially the blue edge, are still uncertain, and significant tension has been found between the theoretical values \citep{dupret05} and the results from \textit{Kepler} for \dsct stars in our Galaxy \citep{murphy19}. 

The right panel of Figure~\ref{fig:CMD_iso_IS} shows a set of BaSTI isochrones with $\rm{[Fe/H]} = -1.0$ dex and a range of ages from 1 to 7~Gyr (in steps of 1~Gyr) consistent with the SMC field in the surroundings of NGC~419. From this Figure and taking account of the high number of \dsct stars outside 2~r$_h$ (see right panel of Figure~\ref{fig:gmos_cmd_variables} and Figure~\ref{fig:finding_chart}), we can infer that the bulk of \dsct stars detected in this work are more likely members of the SMC field, coming from a population between 3-4~Gyr.

\subsection{NGC~419 membership}\label{sec:ngc419}

To further define a membership criterion, we use the structural parameters for the cluster from \citet{ripepi14}. \citet{ripepi14} fit an Elson, Fall, and Freeman (EFF; \citealt{elson87}) profile to their wide-field photometry of the cluster, using a fitting radius of 85\arcsec, beyond which the cluster cannot be distinguished from the background. This radius can be considered as a rough determination of what lies inside and outside the cluster and matches to the size measurement of \citet{bica08}. Additionally, using the \citet{martocchia17} \textit{HST}/ACS photometry we fitted a King profile \citep{king62} to the stellar radial distribution, following the method from \citet{salinas12}. We found a half-light radius (r$_h$) of 45\arcsec, which is consistent with the fitting radius given by \citet{ripepi14}, which is $\sim$ 2~r$_h$. We find that the tidal radius is very sensitive to a chosen limiting~magnitude and is not fully constrained given the limited ACS FOV, and therefore we refrain from giving it. \citet{glatt09} nevertheless gave $r_t\sim 175\arcsec$ based on the number density profile from a similar ACS set of images.

There are only six \dsct stars (V30, V47, V50, V51, V52, and V53) within 85\arcsec\,from the center of NGC~419 (see Figure~\ref{fig:finding_chart} and Figure~\ref{fig:gmos_cmd_variables}, middle panel). The rest 48 \dsct stars are likely members of the SMC field. This is a very low number of \dsct stars, especially comparing with NGC~1846 with $\sim 50$ \dsct star members. However, we are aware of the observational bias we have regarding the detection of \dsct stars in the regions closer to the center due to the crowding. In addition, five out of the six \dsct stars within 2~r$_h$ seem to be clumped in a south-west region (see Figure~\ref{fig:finding_chart}). Since there are no reasons to think that \dsct stars should be clumped in a particular region in the cluster, we believe that this is an indication that we are missing \dsct stars within NGC~419.

There are three factors working in tandem to produce this lack of \dsct stars in the cluster:

\begin{itemize}

\item \textit{Distance}. In general, a cluster significantly more distant than another will obviously imply a larger crowding and an added difficulty to identify variables. This should be only a minor problem in this case since the cluster appears to be significantly in front of the SMC body, even possibly at the same distance as the LMC as proposed by \citet{glatt08}.

\item \textit{Concentration}.  From their ACS photometry, \citet{glatt09} derived a concentration of $c=1.06$, with the caveat that concentration depends on the measurement of the tidal radius which is loosely constrained with the ACS FOV. The cluster would be therefore slightly more concentrated than NGC~1846, for which \citet{goudfrooij09} measured $c=0.79$. A larger concentration implies higher central crowding making difficult the detection of variable stars.

\item \textit{Age}. The most interesting factor could be the age of the cluster.  A hint that the cluster may be slightly younger than, for example, the age measured by \citet{goudfrooij14} of 1.5~Gyr, comes from the fact that there are very few \dsct stars in the MSTO of NGC~419. In addition, those few \dsct stars detected in the inner part of the cluster lie to the red edge of the MSTO (see Figure~\ref{fig:CMD_iso_IS}), instead of scattered across its entire color range. This would happen naturally if only the red part of the MSTO is within the instability strip.
\end{itemize}

\begin{figure*}[t]
    \hspace{-2.5cm}
    \includegraphics[width=1.25\textwidth]{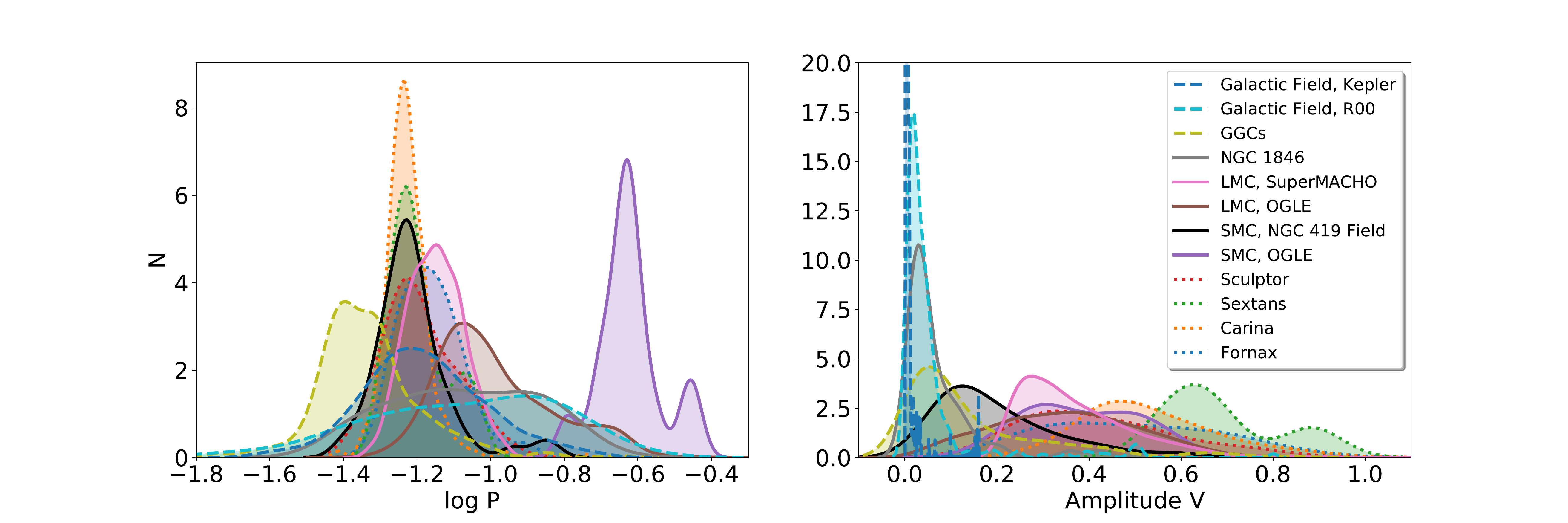}
    \caption{Period (left panel) and amplitude distributions (right panel) of different Galactic and extragalactic \dsct star catalogs. Note that every distribution is normalized so that the area under the curve is one. References for these catalogs plus their mean properties are listed in Table~\ref{tab:catalogs}.}
    \label{fig:Amp_Per_distribution}
\end{figure*}

What are the implications of these results in the context of the extended  MSTO of NGC~419? While in the case of NGC~1846, the contribution of \dsct stars into the broadening of the MSTO is probably very small \citep{salinas18}, in NGC~419 is close to negligible. If the explanation for the extended MSTO lies instead in the existence of a range of populations with different ages \citep{glatt08}, the dearth of \dsct stars in NGC~419, with only a few lying to the red edge of the MSTO, is a natural implication.  The interpretation is more difficult if stellar rotation is instead the cause for the broadening. Stars at the MSTO with significant rotation would appear redder according to the MIST isochrones \citep{choi16}, and even though there is some evidence that indeed the redder stars at NGC~419 MSTO rotate faster \citep{kamann18}, it is unclear whether \dsct stars pulsations are favoured by rotation \citep[e.g.][]{solano97,rasmussen02}.

\subsection{The SMC field \dsct stars population}\label{sec:smc}

\begin{table*}
\caption{Mean properties from the Galactic and extragalactic \dsct star catalogs.}
\label{tab:catalogs}
\hspace{-1cm}
\begin{tabular}{llcccccc}
\toprule
System &  Catalog & N  & $\langle P\rangle$ & $\sigma_P$ & $\langle \Delta_V \rangle$  & $\sigma_{\Delta_V}$ & $\langle$[Fe/H]$\rangle^{(f)}$ \\
& & & (d) & (d) & (mag) & (mag) & (dex) \\
\midrule 
Galactic Field  & \citet{rodriguez00}         & 228$^{(a)}$ & 0.0995 & 0.0538 & 0.060 & 0.107 & $-$0.1 \\
                & \textit{Kepler}; 
                \citet{murphy19}              & 601$^{(a)}$ & 0.0714 & 0.0298 & 0.005 & 0.011 &  $-$0.1 \\
GGCs            & \citet{cohen12}             & 263$^{(b)}$ & 0.0494 & 0.0205 & 0.150 & 0.165 & $-$1.6 \\
NGC~1846        & \citet{salinas18}           & 54 & 0.0933 & 0.0401 & 0.069 & 0.078 & $-$0.5 \\
LMC             & OGLE; \citet{poleski10}     & 2788$^{(c)}$ & 0.1018 & 0.0438 & 0.352 & 0.160 & $-$0.5 \\
                & SuperMACHO; \citet{garg10}  & 1739$^{(d)}$ & 0.0719 & 0.0125 & 0.371 & 0.122 & $-$0.5 \\
SMC             & OGLE; \citet{soszynski02}   & 17 & 0.2433 & 0.0494 & 0.384 & 0.111 & $-$1.0 \\
                & This work (NGC~419 Field)   & 54 & 0.0637 & 0.0195 & 0.199 & 0.127 &  $-$1.0 \\
Sculptor        & \citet{martinezvazquez16}   & 23 & 0.0658 & 0.0148 & 0.422 & 0.152 & $-$1.7 \\
Sextans         & \citet{vivas19}             & 14 & 0.0646 & 0.0119 & 0.689 & 0.119 & $-$1.9 \\
Carina          & \citet{vivas13}$^{(e)}$     & 340 & 0.0613 & 0.0362 & 0.531 & 0.149 & $-$1.7 \\
Fornax          & \citet{poretti08}           & 85 & 0.0702 & 0.0151 & 0.492 & 0.183 &  $-$1.0 \\
\bottomrule
\end{tabular}
\begin{tablenotes}
\item $^{(a)}$Based on the stars selected by \citet{ziaali19}. 
\item $^{(b)}$From a compilation of 28 GGCs with metallicities ranging from [Fe/H] = --2.35 to --0.59 dex.
\item $^{(c)}$We removed those stars that were flagged with a remark in the \citet{poleski10} catalog. The vast majority of them (1488) were uncertain \dsct while 19 were Galactic \dsct stars based on their proper motions and 5 had ``variable mean luminosities''.
\item $^{(d)}$Excluding the 488 stars in common with OGLE \citet{poleski10}.
\item $^{(e)}$\citet{coppola15} report the detection of 435 \dsct stars in Carina but do not give information about the amplitudes.
\item $^{(f)}$References: Galactic Field, \citet{jayasinghe20}; GGCs, \citet{carretta09}; LMC and NGC~1846, \citet{grocholski06, carrera08a}; SMC, \citet{carrera08b,mucciarelli14}, Sextans, \citet{kirby11}, Carina, \citet{koch06}; Sculptor and Fornax, \citet{kirby13}.
\end{tablenotes}
\end{table*}

The \dsct stars population on the SMC has not been purposefully studied, and only a few, mostly serendipitous detections have been made. 

\citet{soszynski02} as part of the OGLE-II campaign in the SMC, found 17 variables with periods between 0.18 and 0.26~d and~magnitudes between $V=18.75$ and 19.82~mag, which they labelled ``probably \dsct stars''. By visual inspection of the individual light curve, using periods reported by OGLE, two of them do not show a reliable light curve\footnote{OGLE005008.48-725916.5 and OGLE005259.56-725605.3}. Therefore, we consider that 15 stars were detected by OGLE in the SMC as ``probably \dsct stars''. Even though the periods are consistent with \dsct stars, their~magnitudes fully overlap with those for RR~Lyrae stars, despite the fact that RR~Lyrae stars are expected to be 1-2~mag brighter than \dsct stars. 

Additionally, \citet{weldrake04}, in a wide-field variability study of the Galactic globular cluster 47~Tuc, found 4 background high-amplitude ($\sim 1$~mag) \dsct stars associated to the SMC, right at the edge of their detection limit.

In this work, we discovered $\sim 50$ \dsct stars as members of the field of the SMC, being the first sizeable detection of \dsct stars ever made in the SMC. Considering that the area covered by GMOS is 30.25~arcmin$^{2}$, the density of the \dsct stars in our field is $\approx 1.6$~\dsct stars/arcmin$^{2}$. This gives us a hint that several thousands of \dsct stars could be found in the body of the SMC, in a survey with similar detection limits and photometric precision than this study. The above estimate also implies that the age distribution of the SMC is homogeneous and similar to the one in the region around NGC 419, which is not necessarily true.

\section{Period and amplitude distributions from different \dsct stars catalogs}\label{sec:period-amplitude}

Figure~\ref{fig:Amp_Per_distribution} shows the period (left panel) and amplitude (right panel) distributions for the 54 \dsct stars in the NGC~419 field (black) plus the period and amplitude distributions of \dsct stars detected so far in different Galactic and extragalactic systems. In order to smooth and better compare the set of distributions, an adaptive kernel density estimator (KDE) was applied over them. We decided to display the logarithm of the period also to make an easily comparison among the distributions. The period distribution of \dsct stars in the field of NGC~419 is quite symmetric with a peak at $\log{P} \sim -1.2$ ($P = 0.06$ d) and extending between $\log{P} \sim -1.5$ ($P = 0.03$ d) and $\log{P} \sim -1.0$ ($P = 0.10$ d). The secondary peak at $\log{P} \sim -0.85$ ($P = 0.14$ d) is an artifact caused by the time span of our observations and those periods are likely aliases. The period distribution of NGC~419 agrees quite well with the period span of other extragalactic systems like Carina \citep{vivas13, coppola15}, Sextans \citep{vivas19}, Sculptor \citep{martinezvazquez16}, Fornax \citep{poretti08}, and the LMC (SuperMACHO, \citealt{garg10}), although the two latter distributions are shifted towards longer periods, with the peak of the distribution at $\log{P} \sim -1.1$ ($P = 0.08$ d). In addition, the \dsct stars in the Galactic field detected by \textit{Kepler} \citep{murphy19} also peaks at $\log{P} \sim -1.0$, displaying a little bit wider distribution than that of NGC~419. However, we obtain a completely different period distribution in this cluster than OGLE-II in the SMC \citep{soszynski02}, which records very long periods. 
This discrepancy is mainly due to the different cadence and photometric depth between OGLE-II (sensitive to periods longer than $\sim 0.2$ d, see \citealt{soszynski02}) and our work (sensitive to periods shorter than $\sim 0.15$ d).

On the other hand, while the \dsct stars of the LMC cluster NGC~1846 \citep{salinas18}, the LMC (OGLE, \citealt{poleski10}), and the Galactic field \citep{rodriguez00} contain larger periods than those observed in NCG~419, the Galactic globular clusters (GGCs) have a distribution that peaks at shorter periods \citep[][and references therein]{cohen12}.

The use of the $V$-band has been widespread in early variability studies. In order to compare the amplitude distribution of the different studies we decided to convert all the amplitudes that were not in $V$ to this band by applying a scaling factor. This scaling factor was calculated by using transforming equations and/or by directly using the ratio of the amplitudes from the catalogs mentioned above. The scaling factors obtained were: $\Delta{V}/\Delta{B} = 0.90$, $\Delta{V}/\Delta{r} = 1.05$, and $\Delta{V}/\Delta{VR} = 1.1$.  We  also use the ratio $\Delta{V}/\Delta{I} = 1.7$ obtained by \citet{salinas16b}. Here the $\Delta$ refers to the amplitude in order to avoid confusion with the extinction coefficients. The amplitude distribution of NGC~419 peaks at 0.15~mag in $V$-band, ranging from 0.05~mag to 0.60~mag.
By inspecting the amplitude distribution of the \dsct stars of the different catalogs, we see that there are different behaviours. It is important noting that the extragalactic catalogs are biased towards larger amplitudes. This is basically due to the limiting~magnitude of the photometry of the studies of these farther galaxies. Therefore, a large number of \dsct stars are not being detected in them because of their low amplitudes, which are smaller than their photometric uncertainties. This is clearly noticeable in the right panel of Figure~\ref{fig:Amp_Per_distribution}. On one hand, systems like Sextans, Carina, Sculptor, and Fornax which have a distance moduli between 19.6 and 20.7~mag \citep{rizzi07,martinezvazquez15,coppola15, vivas19}, show \dsct stars with amplitudes larger than $\sim$0.2~mag. On the other hand, the Galactic field and the Galactic globular clusters have \dsct stars with amplitudes mainly encompassed between 0.001~mag and $\sim$0.2~mag, but also show (to a lesser extent) amplitudes larger than 0.2~mag. In addition, recent studies show that the number of stars with amplitudes of the order of mili-magnitudes is significantly large \citep[e.g.][]{murphy19}. Therefore it is evident that the \dsct stars in the extragalactic fields are biased towards larger amplitudes and we consider that it is not realistic to make a direct comparison among them.

Finally, in Table~\ref{tab:catalogs} we list all the catalogs used in this section together with their mean properties (periods and amplitudes) plus their respective standard deviations. Additionally, we display the mean metallicity for each system in the last column. Excluding the OGLE catalog for the SMC for the reason mentioned above, we can see how the systems with $\rm{[Fe/H]} \le -1.0$ are represented by mean periods ($\overline{\langle P \rangle} = 0.063$ d, $\sigma = 0.007$ d) shorter than those of the systems with $\rm{[Fe/H]} > -1.0$ ($\overline{\langle P \rangle} = 0.088$ d, $\sigma = 0.015$ d).

\begin{table*}
\begin{scriptsize}
\caption{Distance moduli from the \dsct stars.} 
\label{tab:distance}
\hspace{-3.5cm}
\begin{tabular}{ccccccccccccccccccc} 
\toprule
System &  N  & & M11 & & & & CS12 & & & & F15 & & & & MV21 & & & Average \\
\cline{3-5}
\cline{7-9}
\cline{11-13}
\cline{15-17}
       &     &  $\mu_0$ & $\sigma_{\rm rand}$ & $\sigma_{\rm sys}$ & &  $\mu_0$ & $\sigma_{\rm rand}$ & $\sigma_{\rm sys}$ & & $\mu_0$ & $\sigma_{\rm rand}$ & $\sigma_{\rm sys}$ & & $\mu_0$ & $\sigma_{\rm rand}$ & $\sigma_{\rm sys}$ & & $\langle \mu_0 \rangle$ \\
\midrule
NGC~419 & 5 & 19.06 & 0.09 & 0.11 && 18.94 & 0.09 & 0.11 && 18.41 & 0.09 & 0.04 && 18.47 & 0.14 & 0.30 && 18.76 $\pm$ 0.14 \\
SMC Field & 34 & 19.07 & 0.06 & 0.11 && 19.02 & 0.06 & 0.11 && 18.63 & 0.06 & 0.04 && 18.60 & 0.08 & 0.30 && 18.86 $\pm$ 0.11 \\
\bottomrule
\end{tabular}
\begin{tablenotes}
\item \textbf{Notes.} N is the number of stars used in the distance modulus calculation. We exclude those stars with uncertain periods (including double mode and likely alias \dsct stars, see Table~\ref{tab:comments} in the Appendix~\ref{sec:comments}). The distance moduli were obtained using the period-luminosity relation from \citet[][M11]{mcnamara11}, \citet[][CS12]{cohen12}, \citet[][F15]{fiorentino15}, and Mart\'inez-V\'azquez et al. in preparation (MV21).
\end{tablenotes}
\end{scriptsize}
\end{table*}

\section{Distance from \dsct stars}\label{sec:distance}

Like Cepheids and RR~Lyrae stars although fainter and with shorter periods and lower amplitudes, \dsct stars are pulsating variable stars and they can be used as standard candles. As intrinsically fainter stars, precise distances are more difficult to obtain and the construction of its period-luminosity (PL) relations remains a challenge. Despite the difficulties, many PL relations have been explored, both theoretically and empirically \citep[][and Mart\'inez-V\'azquez et al. in preparation]{nemec94,mcnamara11, cohen12,fiorentino15,ziaali19,jayasinghe20}. 

We use the sample of \dsct stars in the field of NGC~419 to estimate the distance -- excluding those \dsct stars with uncertain periods (see Table~\ref{tab:comments} in Appendix~\ref{sec:comments}). Table~\ref{tab:distance} shows the values of the distance modulus calculated from several period-luminosity relations using the \dsct stars of NGC~419 and the field of the SMC (see \S~\ref{sec:ngc419} and \S~\ref{sec:smc}). All these distances are reddening corrected assuming an extinction coefficient of A$_V = 0.25$~mag (see \S~\ref{sec:dsct}).

From the empirical PL relations of \citet{mcnamara11}, assuming a metallicity of $-1.0$ dex and $-0.55$ dex for NGC~419, we obtain a similar distance modulus for both fields ($\mu_0 \sim 19.1$~mag). When using the empirical period-luminosity relation of \citet{cohen12}, we get $\mu_0 = 19.02$~mag for the SMC, but a relatively smaller distance modulus for NGC~419 ($\mu_0 = 18.94$~mag). All these values fit within $0.7 \sigma$. From the theoretical PL relation of \citet[][$Z=0.001$, $Z=0.008$]{fiorentino15}, we get closer distance moduli for the SMC field ($\mu_0 = 18.63$~mag) and for NGC~419 ($\mu_0 = 18.41$~mag). In this case, the difference between the SMC field and NGC~419 is at $1\sigma$. Additionally, using a new PL relation for \dsct stars (Mart\'inez-V\'azquez et al. in preparation) we assess a distance modulus of 18.47~mag for NGC~419 and of 18.60~mag for the SMC field.

The distance moduli obtained using \citet{mcnamara11} and \citet{cohen12} are farther than the distance moduli obtained with \citet{fiorentino15} and Mart\'inez-V\'azquez et al. (in preparation). The dispersion of each distance modulus distribution is 0.4~mag in the SMC field (0.2~mag in NGC~419), which makes all the previous distance moduli obtained compatible considering their systematic uncertainties. In particular, the weighted averaged distance moduli of the previous values are $18.76 \pm 0.14$~mag for NGC~419 and $18.86 \pm 0.11$~mag for the SMC field. This places NGC~419 at 56 kpc, slightly in front of the SMC field (at 59 kpc). However, both measurements fit within $1\sigma$.

As part of the VMC survey, \citet{muraveva18} obtained the distance distribution of the SMC based on RR~Lyrae stars. We identify that our NGC~419 field is in the ``SMC 4\_4'' tile  and has a distribution of distance moduli from 18.4 to 19.4~mag (measured over 250 RR~Lyrae stars), with a mean distance modulus of $18.88\pm0.21$~mag. This value is in agreement with our measure based on \dsct stars. 

The distance of NGC~419 obtained here agrees with the values obtained by \citet{glatt08} and \citet{girardi09}.

\section{Summary and Conclusions}\label{sec:conclusions}

In this paper, we have explored for the first time the population of \dsct stars in the intermediate-age globular cluster NGC~419, and in its surrounding field on the SMC. We have discovered 54 \dsct stars plus three eclipsing binaries on a 5.5\arcmin $\times$ 5.5\arcmin\, area centered on the cluster using Gemini-S/GMOS time series photometry. 

Previous detections of \dsct stars in the SMC only account for less than 20 \dsct stars and did not provide accurate measurements since these stars lie at the~magnitude limit of such studies. 

We have associated six \dsct stars as probable members of the cluster, since they reside within 2~r$_h$ of NGC~419. The remaining 48 \dsct stars, given their position beyond 2~r$_h$ and their location in the CMD, are likely members of the field of the SMC. The latter detections allow us to provide a estimation of the density of \dsct stars in this field of the SMC ($\approx$ 1.6 \dsct stars per arcmin$^{2}$). 
This gives us a clue about the possibility to find tens of thousands of \dsct stars in the field of the SMC with a similar amplitude distribution -- under the assumption that the age distribution of the SMC is homogeneous and similar to the one in the region around NGC 419, which is not necessarily true.

The lack of bright \dsct stars near the MSTO of the cluster made us review different scenarios in terms of crowding and age. Obviously the crowding is an important factor. However, by comparing with NGC~1846 -- which has a significant number of \dsct stars in the MSTO -- the most plausible scenario is that NGC~419 is slightly younger than what the latest age measurements indicate \citep[1.5~Gyr,][]{goudfrooij14,martocchia17}. Thus, the blue part of the MSTO falls outside the instability strip, preventing the pulsation of the stars in that stage. In fact, we performed an isochrone fitting to the ACS color-magnitude diagram of NGC~419 finding that the cluster is well described by a 1.2~Gyr population, with A$_V = 0.25$~mag and a distance modulus of 18.75~mag.

Based on this, the broadening of the MSTO morphology of NGC~419 due to \dsct stars according to the modeling of \citet{salinas16b} is close to negligible. The dearth of \dsct stars in NGC~419, with only a few lying to the red edge of the MSTO, leaves only two possible alternatives for explaining the broadening of the MSTO in the cluster. A range of ages in the stellar population of the cluster would naturally produce a broadening \citep[see e.g.][]{glatt08}; if such ages are young enough, no \dsct stars would be produced in the MSTO but below, as we observe in NGC~419. The second alternative is stellar rotation, whose effects on pulsating stars still remain unclear.

The period distribution of the \dsct stars detected in this work ranges from $0.04 \lesssim P \lesssim$ 0.15 d. In particular, it is very similar to that detected in other Galactic and extragalactic systems. The amplitude distribution ranges from $0.05 \lesssim \Delta r \lesssim 0.60$~mag and it is likely biased. \textit{Kepler} and TESS have detected a significant number of \dsct stars with low amplitudes \citep[see e.g.,][]{murphy19}. Given the distance and data quality, low amplitude \dsct stars cannot be detected in NGC~419.

Finally, we use the \dsct stars discovered in this work to calculate the distance modulus to NGC~419 and the SMC field using different period-luminosity relations. The average distance moduli obtained are $18.76\pm0.14$~mag for NGC~419 and $18.86\pm0.11$ mag for the SMC field. The latter agrees with the distance modulus found in this field by \citet{muraveva18}. The distance of NGC~419 matches with the previous values obtained by \citet{glatt08} and \citet{girardi09}.

Having a census of \dsct stars in the SMC in the future will open a new window to study the intermediate-age population of the galaxy and also to reconstruct the star formation history of the whole body of the SMC by comparing with old and young population tracers already available for this galaxy. Furthermore, increasing the number of \dsct stars in extragalactic systems will help us to better understand the behaviour of these peculiar pulsating stars that so far are the lesser-used and not so well-known of the standard candles.

\acknowledgments
{We thank our anonymous referee for the careful report of the previous version of this manuscript which helped to improve and clarify the paper. 

We thank Silvia Martocchia and Antonino Milone for making their \textit{HST}/ACS photometry of NGC~419 available, and Alessio Mucciarelli for fruitful discussions about the metal content of NGC~419. We also thank Elham Ziaali for useful discussions regarding the Galactic \dsct stars and for sharing his catalog with us, and Simon Murphy for clarifying us some aspects about the \textit{Kepler} \dsct star catalog.

This research has made use of the NASA/IPAC Infrared Science Archive, which is funded by the National Aeronautics and Space Administration and operated by the California Institute of Technology.}

\facilities{Gemini-S, IRSA}
\software{{\sc IRAF}, \citep{tody86, tody93}, {\sc DAOPHOT/ALLSTAR} \citep{stetson87, stetson94}, ISIS \citep[v2.1][]{alard2000}, Astropy \citep{Astropy},  Matplotlib \citep{Matplotlib}}

\clearpage

\bibliography{ngc419}{}
\bibliographystyle{aasjournal}

\clearpage

\appendix

\section{Pulsation parameters}\label{sec:puls-params}
\begin{table*}[!h]
\begin{scriptsize}
\caption{Position and mean pulsation properties of the variables stars detected and discovered in NGC~419} 
\label{tab:var}
\hspace{-3.2cm}
\resizebox{1.2\textwidth}{!}{\begin{tabular}{lccccccclcccccccl}
\toprule
ID &        RA &        Dec & $r_m^{(a)}$  & $g_m^{(a)}$  & $i_m^{(a)}$  & RoMS &     J & Type &    Period &   $\langle r \rangle^{(b)}$ &   $\Delta$r & $\langle g \rangle^{(b)}$ &  $\Delta$g &  $\langle i \rangle^{(b)}$ &  $\Delta$i &  Previous Name \\
 & (deg) & (deg) & (mag) & (mag) & (mag) & & & & (days) & (mag) & (mag) & (mag) & (mag) & (mag) & (mag) & \\
\midrule
  V1 & 16.927417 & -72.847525 &  22.19 &  22.24 &  22.40 &  3.9 &   5.0 &    D &   0.05127: & 22.21 & $>$0.21 & 22.25 & 0.22: & 22.40 & 0.47: &                    \\
  V2 & 16.928590 & -72.883397 &  21.72 &  21.78 &  21.84 &  7.6 &  10.3 &    D &   0.05911  & 21.77 & $>$0.41 & 21.95 & 0.56: & 21.82 & 0.41: &                    \\
  V3 & 16.932629 & -72.884952 &  21.48 &  21.56 &  21.53 &  3.7 &   7.3 &    D &   0.08892: & 21.51 &   0.20: & 21.57 & 0.35: & 21.56 & 0.22: &                    \\
  V4 & 16.936864 & -72.865662 &  18.03 &  17.87 &  18.36 & 16.9 &  33.0 &    E &   2.16870  &  ...  &   ...   &  ...  &  ...  & ...   & ...   &  OGLE-SMC-ECL-5297 \\
  V5 & 16.939052 & -72.912768 &  21.51 &  21.70 &  21.53 &  2.9 &   3.9 &    D &   0.05677  & 21.50 & $>$0.10 & 21.67 & 0.20: & 21.53 & 0.09: &                    \\
  V6 & 16.939722 & -72.872235 &  21.72 &  21.95 &  21.77 &  2.8 &   3.7 &    D &   0.07459: & 21.72 &    0.12 & 21.95 & 0.12: & 21.75 & 0.18: &                    \\
  V7 & 16.942151 & -72.890242 &  21.40 &  21.61 &  21.44 &  2.3 &   2.8 &    D &   0.06611  & 21.40 &    0.06 & 21.63 & 0.10: & 21.44 & 0.13: &                    \\
  V8 & 16.943567 & -72.927549 &  21.85 &  22.02 &  21.89 &  2.1 &   3.0 &    D &   0.04955: & 21.85 &   0.08: & 22.01 & 0.11: & 21.89 & 0.07: &                    \\
  V9 & 16.959281 & -72.907626 &  21.39 &  21.64 &  21.39 &  5.0 &   7.4 &    D &   0.07462  & 21.40 &    0.16 & 21.63 & 0.22: & 21.41 & 0.15: &                    \\
 V10 & 16.959599 & -72.928604 &  21.76 &  21.90 &  21.83 &  4.0 &   5.3 &    D &   0.05885  & 21.78 &    0.16 & 21.91 & 0.30: & 21.85 & 0.27: &                    \\
 V11 & 16.960351 & -72.847973 &  16.21 &  16.63 &  16.14 &  6.5 &   8.9 &    C &   2.70441  &  ...  &   ...   &  ...  &  ...  & ...   & ...   &  OGLE-SMC-CEP-3982 \\
 V12 & 16.960539 & -72.914231 &  22.12 &  22.30 &  22.19 &  2.5 &   3.2 &    D &   0.06177  & 22.12 &    0.14 & 22.33 & 0.27: & 22.17 & 0.16: &                    \\
 V13 & 16.961316 & -72.916217 &  21.48 &  21.63 &  21.57 &  6.5 &  10.7 &    D &   0.05490  & 21.48 &    0.35 & 21.67 & 0.41: & 21.55 & 0.12: &                    \\
 V14 & 16.962542 & -72.898692 &  22.17 &  22.39 &  22.23 &  2.4 &   3.1 &    D &   0.06699: & 22.17 & $>$0.09 & 22.39 & 0.10: & 22.23 & 0.01: &                    \\
 V15 & 16.965802 & -72.879215 &  17.48 &  17.30 &  17.82 &  2.2 &   2.8 &    E &  60.25810  &  ...  &   ...   &  ...  &  ...  & ...   & ...   &  OGLE-SMC-ECL-5317 \\
 V16 & 16.966121 & -72.894951 &  18.99 &  18.75 &  19.30 & 84.6 & 122.1 &    E &   0.90925  & 18.83 &    0.69 & 18.65 & 0.74: & 19.06 & 0.77: &  OGLE-SMC-ECL-5318 \\
 V17 & 16.967447 & -72.891335 &  22.13 &  22.33 &  22.10 &  2.5 &   3.5 &    D &   0.06132  & 22.14 &    0.14 & 22.34 & 0.19: & 22.16 & 0.21: &                    \\
 V18 & 16.967564 & -72.898786 &  21.48 &  21.63 &  21.55 &  6.0 &   7.2 &    D &   0.06187  & 21.47 &    0.33 & 21.54 & 0.58: & 21.54 & 0.21: &                    \\
 V19 & 16.967748 & -72.909831 &  21.31 &  21.48 &  21.38 &  4.5 &   5.9 &    D &   0.06598  & 21.31 &    0.12 & 21.49 & 0.15: & 21.38 & 0.07: &                    \\
 V20 & 16.969233 & -72.890059 &  22.20 &  22.36 &  22.31 &  1.8 &   2.4 &    D &   0.04238: & 22.20 & $>$0.08 & 22.36 & 0.10: & 22.30 & 0.00: &                    \\
 V21 & 16.970038 & -72.839723 &  22.05 &  22.16 &  22.13 &  3.3 &   4.8 &    D &   0.05303: & 22.05 & $>$0.22 & 22.15 & 0.31: & 22.13 & 0.13: &                    \\
 V22 & 16.971265 & -72.849554 &  21.46 &  21.62 &  21.53 &  5.6 &   7.3 &    D &   0.08065  & 21.46 &    0.27 & 21.67 & 0.44: & 21.48 & 0.28: &                    \\
 V23 & 16.974041 & -72.885406 &  22.16 &  22.36 &  22.23 &  2.7 &   3.3 &    D &   0.04324  & 22.17 &    0.13 & 22.37 & 0.19: & 22.23 & 0.17: &                    \\
 V24 & 16.974785 & -72.919927 &  21.92 &  22.08 &  21.96 &  3.8 &   4.3 &    D &   0.06257  & 21.91 & $>$0.26 & 22.07 & 0.31: & 21.93 & 0.15: &                    \\
 V25 & 16.984115 & -72.893033 &  21.54 &  21.69 &  21.61 &  1.9 &   2.2 &    D &   0.05750: & 21.54 & $>$0.05 & 21.70 & 0.06: & 21.62 & 0.13: &                    \\
 V26 & 17.011986 & -72.918993 &  21.76 &  21.96 &  21.82 &  1.9 &   2.3 &    D &   0.06111: & 21.77 & $>$0.08 & 21.98 & 0.17: & 21.82 & 0.16: &                    \\
 V27 & 17.012055 & -72.906415 &  21.91 &  22.06 &  21.99 &  2.8 &   3.6 &    D &   0.05994  & 21.91 & $>$0.18 & 22.08 & 0.20: & 22.00 & 0.23: &                    \\
 V28 & 17.013551 & -72.851505 &  22.15 &  22.30 &  22.27 &  3.1 &   4.0 &    D &   0.05382  & 22.17 &    0.28 & 22.30 & 0.28: & 22.26 & 0.27: &                    \\
 V29 & 17.013617 & -72.925285 &  21.32 &  21.55 &  21.37 &  2.1 &   2.8 &    D &   0.06727  & 21.33 &    0.08 & 21.56 & 0.10: & 21.36 & 0.00: &                    \\
 V30* & 17.014896 & -72.881626 &  21.86 &  22.00 &  21.97 &  2.0 &   2.4 &    D &   0.14703: & 21.84 &   0.09: & 21.95 & 0.29: & 21.90 & 0.57: &                    \\
 V31 & 17.022893 & -72.923688 &  21.87 &  21.93 &  21.96 &  6.3 &   7.4 &    D &   0.06328  & 21.91 &    0.56 & 22.07 & 0.78: & 21.95 & 0.44: &                    \\
 V32 & 17.023467 & -72.912373 &  21.67 &  21.92 &  21.72 &  4.1 &   5.0 &    D &   0.06490  & 21.69 & $>$0.24 & 21.89 & 0.34: & 21.73 & 0.17: &                    \\
 V33 & 17.030176 & -72.927587 &  17.31 &  17.66 &  17.30 & 21.8 &  21.1 &    C &   1.09662  &  ...  &   ...   &  ...  &  ...  & ...   & ...   &  OGLE-SMC-CEP-4008 \\
 V34 & 17.030272 & -72.917776 &  21.37 &  21.35 &  21.60 &  2.9 &   4.0 &    E &   0.45498  & 21.40 &    0.33 & 21.36 & 0.46: & 21.57 & 0.27: &                    \\
 V35 & 17.049675 & -72.849494 &  21.65 &  21.67 &  21.71 &  3.8 &   3.8 &    D &   0.07714  & 21.69 &    0.32 & 21.77 & 0.48: & 21.69 & 0.28: &                    \\
 V36 & 17.050798 & -72.929004 &  20.21 &  20.28 &  20.32 & 12.8 &  19.2 &    E &   2.32527: & 20.25: &  0.76: & 20.38: & 1.45: & 20.36: & 0.80: &                    \\
 V37 & 17.051782 & -72.908345 &  21.82 &  22.01 &  21.90 &  1.6 &   2.2 &    D &   0.04826: & 21.82 &   0.07: & 22.02 & 0.17: & 21.90 & 0.05: &                    \\
 V38 & 17.052973 & -72.851870 &  21.90 &  22.03 &  22.01 &  2.9 &   3.9 &    D &   0.05353  & 21.91 &    0.19 & 22.05 & 0.22: & 22.01 & 0.11: &                    \\
 V39 & 17.055457 & -72.919961 &  21.58 &  21.62 &  21.74 &  4.5 &   4.5 &    D &   0.04859  & 21.61 &    0.12 & 21.88 & 0.01: & 21.85 & 0.05: &                    \\
 V40 & 17.069977 & -72.838887 &  17.75 &  17.78 &  17.80 & 24.7 &  42.4 &    C &   0.78835  &  ...  &   ...   &  ...  &  ...  & ...   & ...   &  OGLE-SMC-CEP-4027 \\
 V41 & 17.071523 & -72.885379 &  17.36 &  17.73 &  17.32 &  7.1 &  12.7 &    C &   1.04282  &  ...  &   ...   &  ...  &  ...  & ...   & ...   &  OGLE-SMC-CEP-4028 \\
 V42 & 17.080895 & -72.879923 &  17.89 &  18.29 &  17.84 &  3.8 &   4.4 &    C &   0.80029  &  ...  &   ...   &  ...  &  ...  & ...   & ...   &  OGLE-SMC-CEP-4032 \\
 V43 & 17.086087 & -72.879071 &  18.10 &  18.86 &  17.85 &  4.8 &   4.9 &    E & 250.23399  &  ...  &   ...   &  ...  &  ...  & ...   & ...   &  OGLE-SMC-ECL-7769 \\
 V44 & 17.099490 & -72.848562 &  22.35 &  22.63 &  22.33 &  4.0 &   6.1 &    D &   0.13588  & 22.48 &    0.54 & 22.70 & 0.55: & 22.52 & 0.65: &                    \\
 V45 & 17.099514 & -72.850945 &  22.07 &  22.10 &  22.08 &  4.6 &   6.4 &    D &   0.04921  & 22.11 &    0.41 & 22.23 & 0.72: & 22.14 & 0.42: &                    \\
 V46 & 17.103834 & -72.852755 &  20.64 &  20.92 &  20.62 &  4.3 &   5.8 &    D &   0.05742  & 20.64 &    0.15 & 20.93 & 0.21: & 20.54 & 0.41: &                    \\
 V47* & 17.105888 & -72.894902 &  21.57 &  21.80 &  21.63 &  4.2 &   4.7 &    D &   0.06225: & 21.57 & $>$0.13 & 21.81 & 0.15: & 21.69 & 0.30: &                    \\
 V48 & 17.105922 & -72.889339 &  20.06 &  19.97 &  20.28 &  2.9 &   4.2 &    E &   2.70531  &  ...  &   ...   &  ...  &  ...  & ...   & ...   &  OGLE-SMC-ECL-5389 \\
 V49 & 17.106071 & -72.864994 &  16.25 &  16.71 &  16.14 & 62.9 &  59.5 &    C &   1.42403  &  ...  &   ...   &  ...  &  ...  & ...   & ...   &  OGLE-SMC-CEP-4037 \\
 V50* & 17.110899 & -72.894365 &  21.79 &  21.90 &  21.86 &  2.6 &   3.0 &    D &   0.05671: & 21.80 &   0.12: & 21.91 & 0.13: & 21.92 & 0.31: &                    \\
 V51* & 17.115403 & -72.893292 &  21.36 &  21.46 &  21.50 &  2.1 &   2.4 &    D &   0.05186: & 21.36 &   0.05: & 21.46 & 0.04: & 21.51 & 0.21: &                    \\
 V52* & 17.124513 & -72.891063 &  21.52 &  21.71 &  21.65 &  2.1 &   2.4 &    D &   0.05751: & 21.52 &   0.07: & 21.70 & 0.06: & 21.66 & 0.19: &                    \\
 V53* & 17.126239 & -72.891730 &  21.57 &  21.77 &  21.66 &  2.6 &   2.9 &    D &   0.06106: & 21.58 &   0.08: & 21.79 & 0.08: & 21.68 & 0.24: &                    \\
 V54 & 17.126670 & -72.902176 &  17.57 &  17.89 &  17.58 & 42.6 &  65.8 &    C &   0.91508  &  ...  &   ...   &  ...  &  ...  & ...   & ...   &  OGLE-SMC-CEP-4045 \\
 V55 & 17.128872 & -72.845712 &  21.91 &  22.00 &  22.15 &  4.2 &   4.2 &    D &   0.05753  & 22.11 & $>$0.22 & 22.38 & 0.40: & 22.07 & 0.26: &                    \\
 V56 & 17.160577 & -72.891848 &  21.21 &  21.47 &  21.20 &  3.9 &   5.8 &    D &   0.05677: & 21.21 &    0.10 & 21.48 & 0.17: & 21.19 & 0.16: &                    \\
 V57 & 17.164492 & -72.911477 &  21.85 &  22.09 &  21.89 &  5.7 &   7.1 &    D &   0.06624  & 21.86 &    0.26 & 22.02 & 0.47: & 21.91 & 0.20: &                    \\
 V58 & 17.170817 & -72.890989 &  21.47 &  21.64 &  21.48 & 10.9 &  15.0 &    D &   0.06565  & 21.50 &    0.42 & 21.70 & 0.61: & 21.58 & 0.39: &                    \\
 V59 & 17.171052 & -72.923987 &  21.94 &  22.07 &  22.06 &  2.8 &   3.8 &    D &   0.03899  & 21.94 &    0.11 & 22.06 & 0.12: & 22.05 & 0.19: &                    \\
 V60 & 17.175089 & -72.856043 &  16.48 &  16.73 &  16.46 & 17.7 &  22.8 &    C &   1.77642  &  ...  &   ...   &  ...  &  ...  & ...   & ...   &  OGLE-SMC-CEP-4055 \\
 V61 & 17.178552 & -72.924637 &  21.66 &  21.78 &  21.72 &  7.3 &  10.6 &    D &   0.05895  & 21.70 &    0.34 & 21.80 & 0.53: & 21.76 & 0.36: &                    \\
 V62 & 17.186289 & -72.894658 &  16.91 &  17.34 &  16.86 & 72.4 &  67.3 &    C &   1.83185  &  ...  &   ...   &  ...  &  ...  & ...   & ...   &  OGLE-SMC-CEP-4062 \\
 V63 & 17.191218 & -72.914169 &  21.30 &  21.45 &  21.35 &  3.1 &   5.2 &    D &   0.05351: & 21.30 &   0.09: & 21.45 & 0.15: & 21.34 & 0.00: &                    \\
 V64 & 17.197690 & -72.891658 &  22.37 &  22.55 &  22.43 &  5.0 &   6.0 &    D &   0.07581: & 22.37 &   0.20: & 22.57 & 0.15: & 22.42 & 0.27: &                    \\
 V65 & 17.201044 & -72.893491 &  21.13 &  21.23 &  21.20 &  5.5 &   7.2 &    D &   0.04982  & 21.11 &    0.18 & 21.23 & 0.21: & 21.20 & 0.00: &                    \\
 V66 & 17.203056 & -72.860402 &  22.03 &  22.24 &  22.06 &  2.8 &   3.9 &    D &   0.07178: & 22.02 & $>$0.15 & 22.23 & 0.30: & 22.05 & 0.21: &                    \\
 V67 & 17.208281 & -72.861341 &  22.26 &  22.37 &  22.27 &  3.8 &   4.8 &    D &   0.05729  & 22.25 &    0.28 & 22.39 & 0.30: & 22.27 & 0.14: &                    \\
 V68 & 17.208568 & -72.871081 &  19.21 &  19.05 &  19.42 &  8.0 &  11.1 &    E &   2.17219  &   ... &   ...   &  ...  &  ...  & ...   & ...   &  OGLE-SMC-ECL-5433 \\
 V69 & 17.210417 & -72.842314 &  21.91 &  21.89 &  22.03 &  3.3 &   5.2 &    D &   0.11439: & 21.92 &   0.18: & 21.90 & 0.13: & 22.03 & 0.16: &                    \\
 V70 & 17.216029 & -72.915151 &  16.07 &  16.39 &  16.07 &  9.1 &   9.0 &    C &   2.01063  &  ...  &   ...   &  ...  &  ...  & ...   & ...   &  OGLE-SMC-CEP-4068 \\
 V71 & 17.219944 & -72.885980 &  22.14 &  22.58 &  22.02 &  2.6 &   3.4 &    E &   0.13620  & 22.15 & $>$0.12 & 22.62 & 0.20: & 22.05 & 0.19: &                    \\
 V72 & 17.222930 & -72.898221 &  22.05 &  22.23 &  22.05 &  3.0 &   4.2 &    D &   0.04703  & 22.05 & $>$0.11 & 22.25 & 0.18: & 22.05 & 0.02: &                    \\
 V73 & 17.227135 & -72.916692 &  21.15 &  21.36 &  21.20 &  4.1 &   5.4 &    D &   0.05712  & 21.15 &    0.09 & 21.36 & 0.12: & 21.20 & 0.03: &                    \\
\bottomrule
\end{tabular}
}
\begin{tablenotes}
\item \textbf{Notes.} The classification (column \textit{Type}) is C--Cepheid, D--$\delta$~Scuti, E--Eclipsing binary. Comments on individual variables are in appendix section. $\delta$~Scuti stars within 2~r$_h$ of NGC~419 are those marked with an asterisk in their names. For the OGLE variables we consider the periods published by OGLE. Uncertain periods and amplitudes are indicated with a colon. The $>$ indicates lower limits in the amplitudes. Because of the low number of epochs, we consider the amplitudes in $g$ and $i$ uncertain.
\item $^{(a)}$ These magnitudes are the weighted intensity-average values calculated from the individual measurements.
\item $^{(b)}$ These are the intensity-averaged magnitudes obtained by fitting the light curves with the \cite{layden98}'s templates.
\end{tablenotes}
\end{scriptsize}
\end{table*}

Table~\ref{tab:var} gives the positions, periods, intensity-weighted~magnitudes and classifications for the 73 variable stars detected in this work. 

\clearpage

\section{Time-series data}\label{sec:time-series}

Table~\ref{tab:photometry} publishes the individual $r,g,i$ measurements for the variables detected in this work. Table~\ref{tab:photometry} is published in its entirety in the machine-readable format. A portion is shown here for guidance regarding its form and content.

\begin{table*}[!h]
\begin{scriptsize}
\centering
\caption{Photometry of the variable stars in NGC~419}
\label{tab:photometry}
\begin{tabular}{ccccccccc} 
\toprule
MJD$_r^{(a)}$    &    $r$    &    $\sigma_r$    &    MJD$_g^{(a)}$    &    $g$    &    $\sigma_g$    &    MJD$_i^{(a)}$    &    $i$    &    $\sigma_i$   \\
\midrule
\multicolumn{9}{c}{V1} \\
\hline
   58010.2647  &   22.052  &    0.017  &   58010.2616  &   22.071  &    0.017  &   58010.2731  &   22.228  &    0.037 \\
   58023.0347  &   22.066  &    0.026  &   58023.0415  &   22.122  &    0.048  &   58023.0324  &   22.257  &    0.037 \\
   58023.0400  &   22.111  &    0.035  &   58018.2036  &   22.300  &    0.016  &   58018.2019  &   22.455  &    0.051 \\
   58023.0311  &   22.087  &    0.022  &   58018.2393  &   22.303  &    0.016  &   58018.2508  &   22.568  &    0.060 \\
   58023.0337  &   22.061  &    0.022  &   58018.2752  &   22.221  &    0.012  &   58018.2867  &   22.517  &    0.061 \\
   58023.0301  &   22.086  &    0.021  &   58018.2163  &   22.175  &    0.016  &   58018.2278  &   22.866  &    0.112 \\
   58023.0389  &   22.042  &    0.021  &   58018.1904  &   22.380  &    0.020  &   58010.2961  &   22.662  &    0.056 \\
   58023.0286  &   22.138  &    0.019  &   58018.2624  &   22.194  &    0.017  &            &              &         \\
   58023.0239  &   22.217  &    0.029  &   58010.2846  &   22.239  &    0.027  &            &              &         \\
   ...         &    ...    &    ...    &    ...        &    ...    &     ...   &    ...     &    ...       &     ... \\
\bottomrule
\end{tabular}
\begin{tablenotes}
\item \textbf{Notes.} Table~\ref{tab:photometry} is published in its entirety in the machine-readable format. A portion is shown here for guidance regarding its form and content.
\item $^{(a)}$Modified Julian Date at mid-exposure.
\end{tablenotes}
\end{scriptsize}
\end{table*}

\clearpage

\section{Light curves of the \dsct stars variables}\label{sec:lcv_dsct}

\begin{figure*}[!h]
    \centering
    \includegraphics[width=0.7\textwidth]{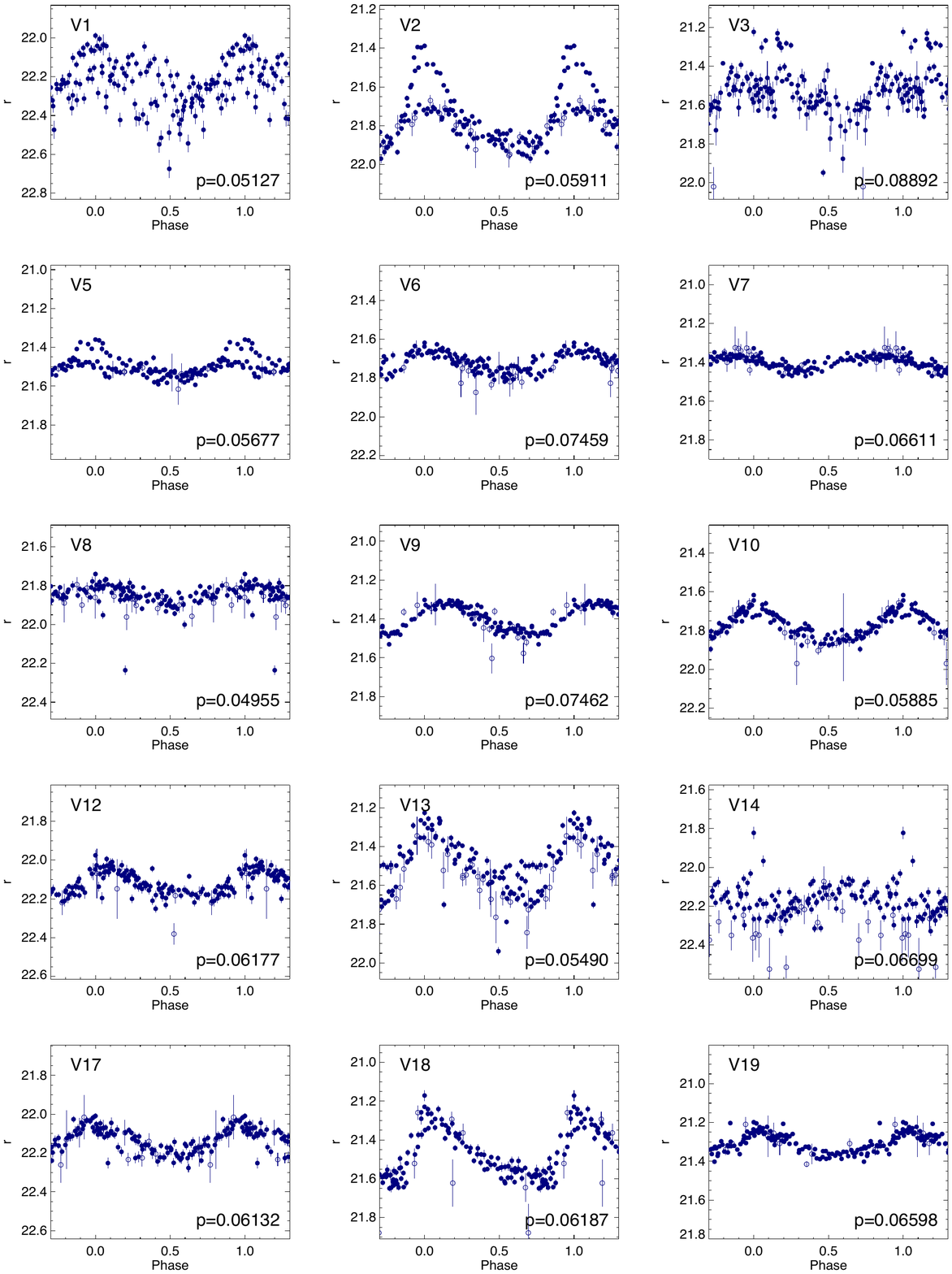}
    \caption{Light curves in the $r$ band of the \dsct stars discovered in this work. Periods (in days) are given in the lower right corner, while the name of the variable is displayed at the top left of each panel. Open symbols show the data for which the uncertainties are larger than 3$\sigma$ above the mean error of a given star and, therefore, they were not considered in the period, amplitude and mean~magnitude calculations.}
    \label{fig:dSct_lcv}
\end{figure*}

Figure~\ref{fig:dSct_lcv} shows the folded $r$-band light curves of the 54 \dsct stars discovered in this work. 

\begin{Contfigure}[!h]
   \centering
    \includegraphics[width=0.7\textwidth]{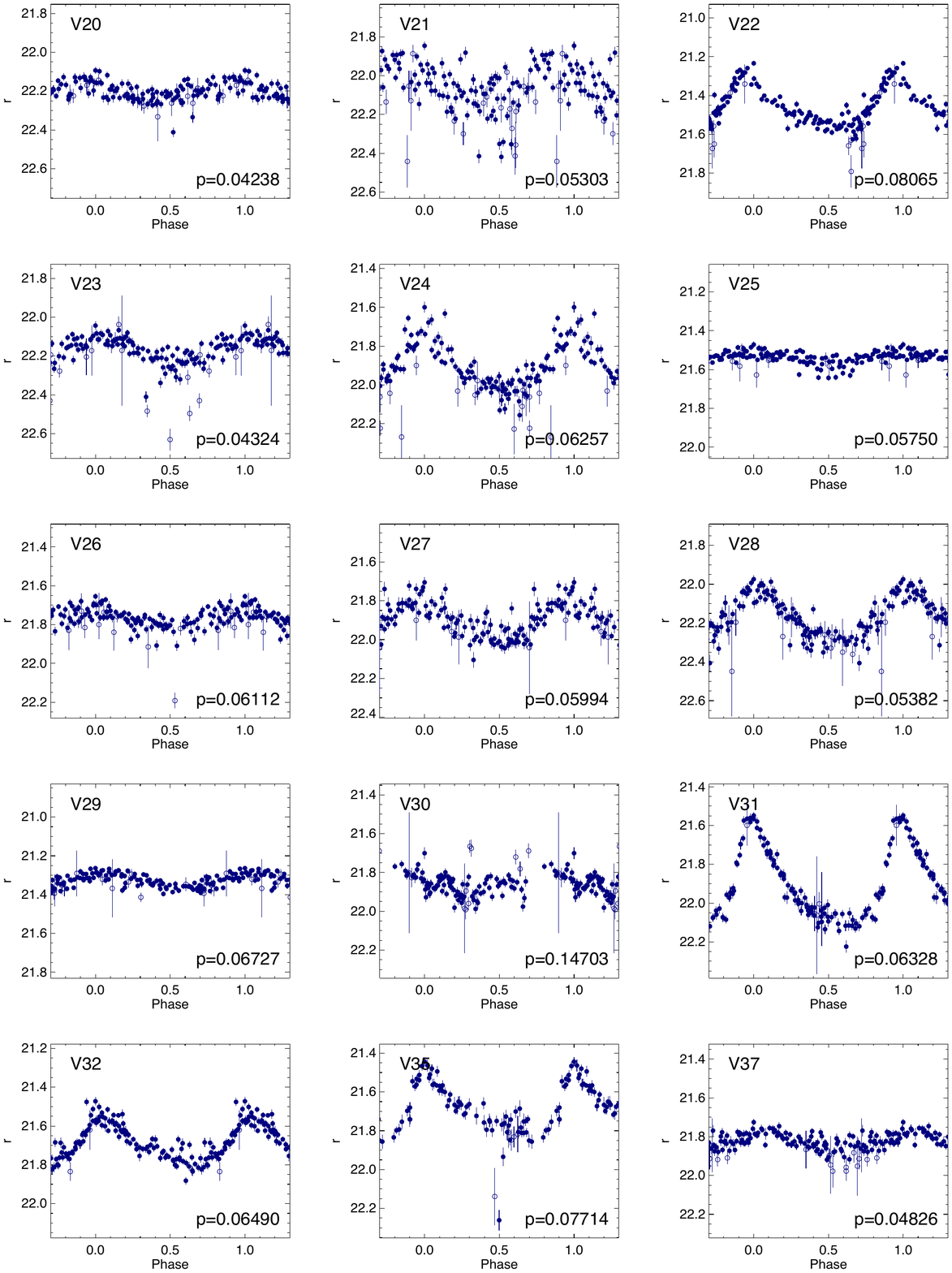}
    \caption{Light curves in the $r$ band of the \dsct stars discovered in this work. Periods (in days) are given in the lower right corner, while the name of the variable is displayed at the top left of each panel. Open symbols show the data for which the uncertainties are larger than 3$\sigma$ above the mean error of a given star and, therefore, they were not considered in the period, amplitude and mean~magnitude calculations.}
    \label{fig:dSct_lcv1}
\end{Contfigure}

\begin{Contfigure}[!h]
   \centering
    \includegraphics[width=0.7\textwidth]{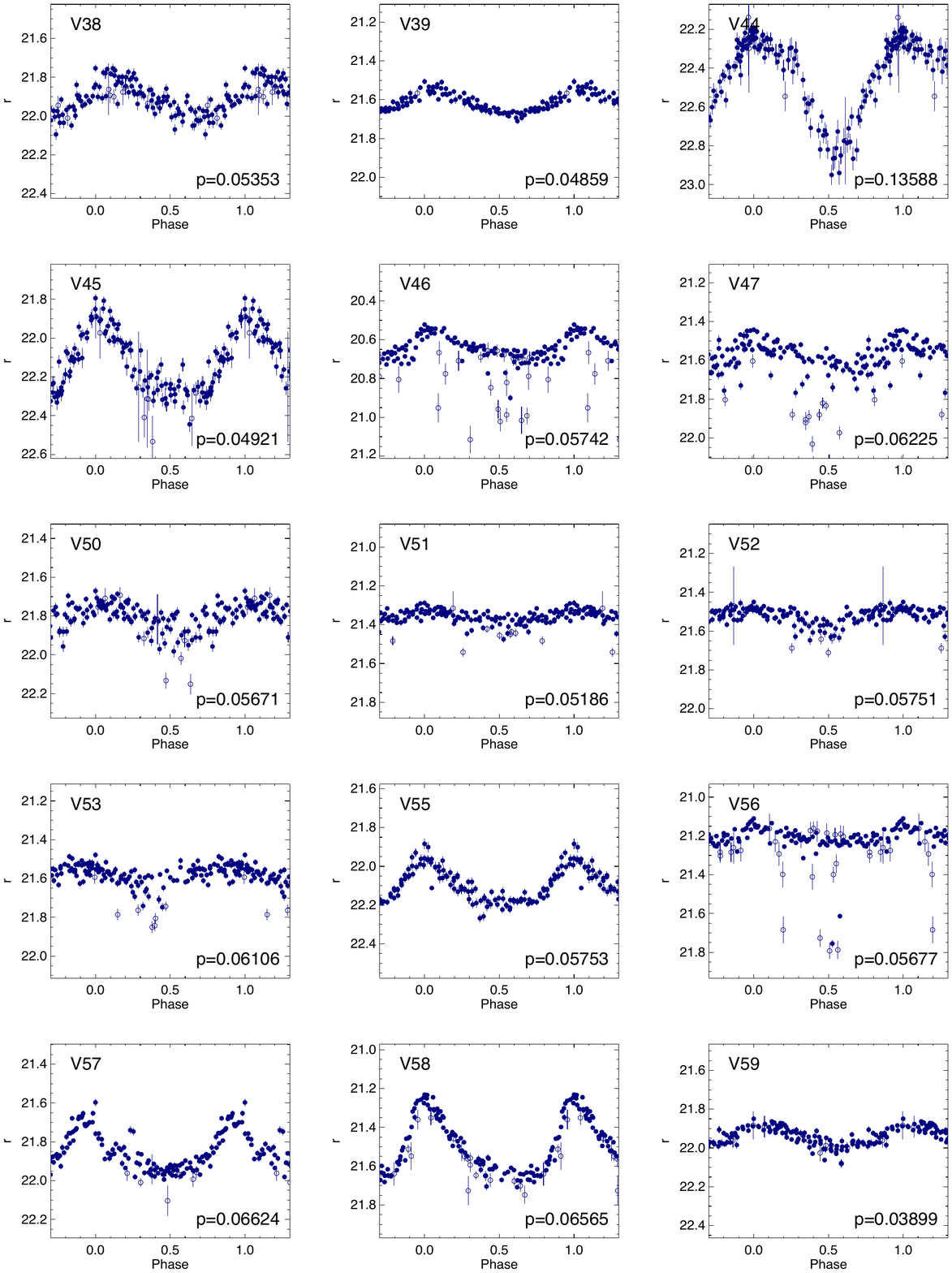}
    \caption{Light curves in the $r$ band of the \dsct stars discovered in this work. Periods (in days) are given in the lower right corner, while the name of the variable is displayed at the top left of each panel. Open symbols show the data for which the uncertainties are larger than 3$\sigma$ above the mean error of a given star and, therefore, they were not considered in the period, amplitude and mean~magnitude calculations.}
    \label{fig:dSct_lcv2}
\end{Contfigure}

\begin{Contfigure}[!h]
   \centering
    \includegraphics[width=0.7\textwidth]{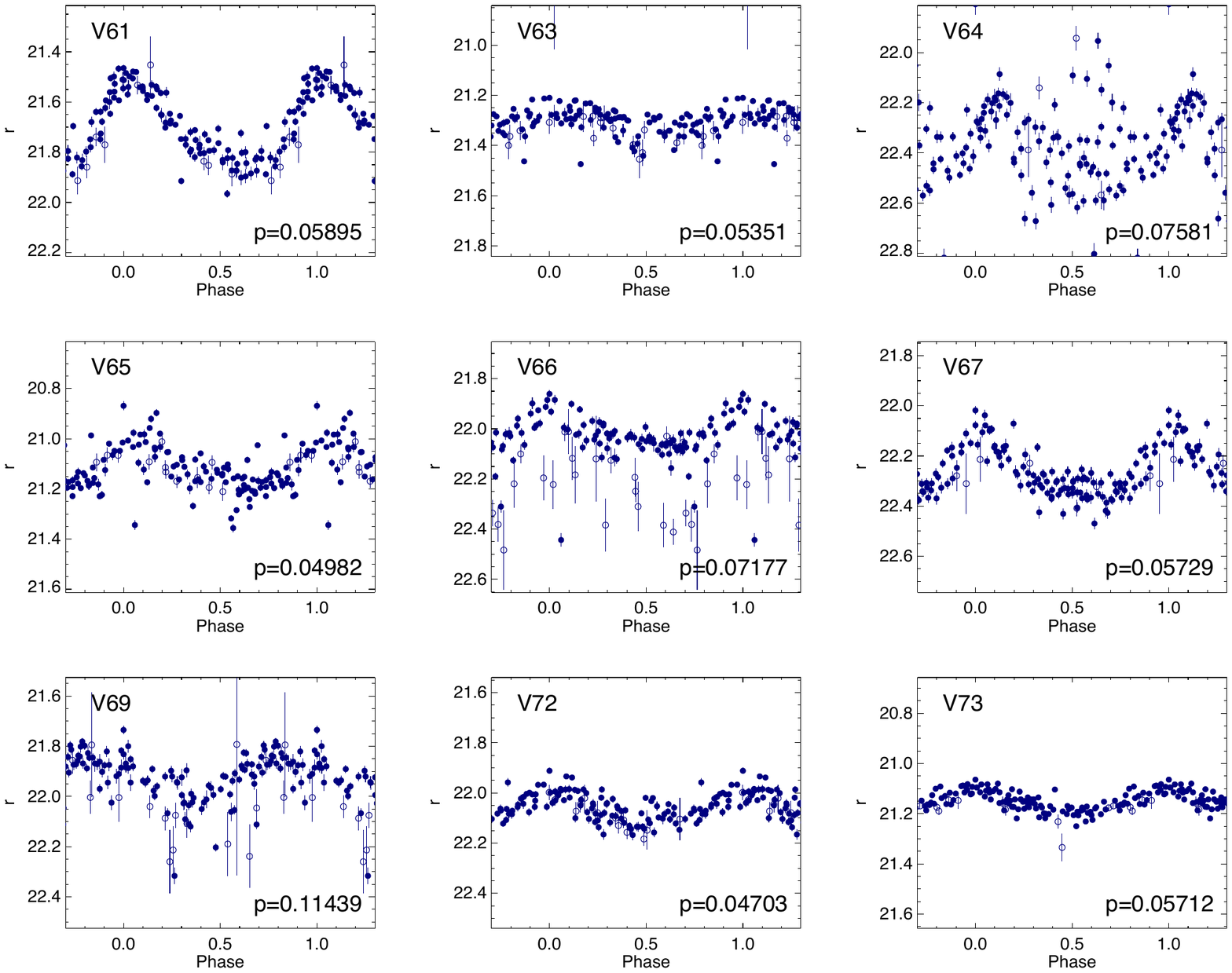}
    \caption{Light curves in the $r$ band of the \dsct stars discovered in this work. Periods (in days) are given in the lower right corner, while the name of the variable is displayed at the top left of each panel. Open symbols show the data for which the uncertainties are larger than 3$\sigma$ above the mean error of a given star and, therefore, they were not considered in the period, amplitude and mean~magnitude calculations.}
    \label{fig:dSct_lcv3}
\end{Contfigure}

\clearpage

\section{Light curves of the best Eclipsing Binaries detected}\label{sec:lcv_ebin}

Figure~\ref{fig:dSct_lcv} shows the folded $r$-band light curves of one OGLE eclipsing binary that we were able to reproduce correctly with our data plus the three eclipsing binaries discovered in this work. 

\begin{figure*}[!h]
    \centering
    \includegraphics[width=0.6\textwidth]{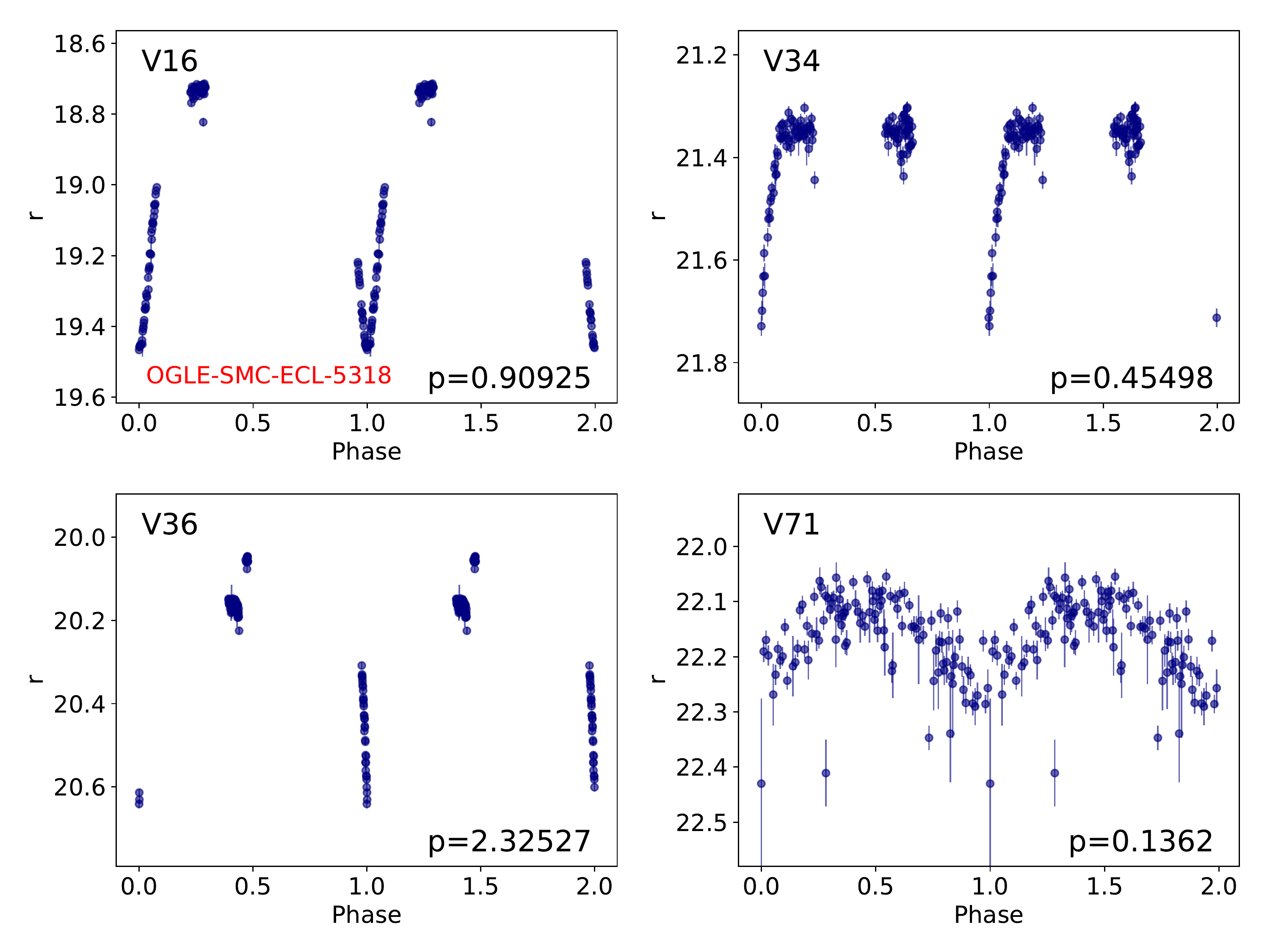}
    \caption{Light curves in the $r$ band of the best eclipsing binaries detected in our data, the OGLE-SMC-ECL-5318 (named as V16) plus the three new eclipsing binary stars discovered.}
    \label{fig:bin_lcv}
\end{figure*}

\clearpage

\section{Comments on individual variable stars}\label{sec:comments}

Table~\ref{tab:comments} records the notes on some individual variables.  
\begin{table*}[!h]
\begin{scriptsize}
\caption{Individual comments on the variables stars detected and discovered in the field of NGC~419.} 
\label{tab:comments}
\hspace{-1cm}
\resizebox{1.1\textwidth}{!}{\begin{tabular}{ccl} 
\toprule
ID  &  OGLE ID          &  Comments                                                                 \\
\midrule
V1    &              	      &  Probable but noisy. Uncertain period. \\
V2    &              	      &  Double mode \dsct star. \\
V3    &             	      &  Noisy. Double mode \dsct star? \\
V4    &  OGLE-SMC-ECL-5297	  &  Eclipsing binary in OGLE but no reliable light curve from our data. \\
V5    &              	      &  Double mode \dsct star. \\
V13   &               	      &  Strange behaviour at minimum light. \\
V14   &              	      &  Noisy. It is in a crowded region. \\
V15   &  OGLE-SMC-ECL-5317	  &  Eclipsing binary in OGLE with long period. It does not show variability in our data.  \\
V16   &  OGLE-SMC-ECL-5318	  &  Most obvious OGLE detection in our data. \\
V20   &              	      &  Noisy. Uncertain period. \\
V21   &             	      &  Noisy. Uncertain period. \\
V24   &              	      &  Period of 0.0567730 d is also possible. \\
V25   &              	      &  Noisy. \\
V26   &              	      &  Noisy. \\
V27   &       	              &  Presence of a few outliers. The star is in a crowded region. \\ 
V30   &              	      &  Noisy. Subluminous. Its period is likely an alias. \\ 
V34   &       		          &  Eclipsing binary with uncertain period. \\
V36   &       		          &  Probable eclipsing binary. \\ 
V37   &              	      &  Noisy. \\
V38   &               	      &  Noisy. Double mode \dsct star? Period of 0.0534902 d is also possible. \\
V39   &                       &  \dsct recovered from ISIS. \\
V40   &  OGLE-SMC-CEP-4027	  &  Close to the edge of the image (probably not good photometry). \\
V43   &  OGLE-SMC-ECL-7769	  &  Eclipsing binary in OGLE with long period. It does not show variability in our data. \\
V44   &       	              &  Close to a bad column (probably not good photometry). Subluminous. Its period is likely an alias. \\
V48   &  OGLE-SMC-ECL-5389	  & Eclipsing binary in OGLE but it does not show variability in our data. \\      
V50   &               	      &  Noisy. \\
V51   &              	      &  Noisy. \\
V53   &              	      &  Noisy. \\
V55   &                       & \dsct star recovered from ISIS. \\
V64   &              	      &  Noisy with some outliers. Double mode \dsct star? Uncertain period. \\
V66   &              	      &  Noisy with some outliers. Double mode \dsct star? \\
V68   &  OGLE-SMC-ECL-5433    &  Eclipsing binary in OGLE but it does not show variability in our data. \\
V69   &              	      &  Subluminous. Its period is likely an alias. \\    
V71   &              	      &  Variable with a reddish color. Probable eclipsing binary. \\
\bottomrule
\end{tabular}}
\end{scriptsize}
\end{table*}

\end{document}